\documentclass[prb,amssymb,aps]{revtex4}
\usepackage{graphicx}
\usepackage{mathptm, amssymb}

\begin{document}

\title{Levy defects in fluctuating pattern of liquids. A quasi thermodynamic
approach to the dynamic glass transition in classical molecular liquids.}
\author{E. Donth}
\address{Institut f\"ur Physik, Universit\"at Halle, D-06120 Halle (Saale),
Germany\\
electronic mail: donth@physik.uni-halle.de}

\begin{abstract}
This theoretical paper is to advance a phenomenological, quasi thermodynamic
approach to the dynamics of classical liquids which uses the Levy
distribution of probability theory. Doubts from the chemical physics
community about the application of its unusual properties to this field are
tried to be removed. In particular, to understand the preponderant component
of the Levy sum for Glarum Levy defects and Fischer speckles, the classical
mathematical proof [D. A. Darling, Trans. Amer. Math. Soc. {\bf 73}, 95
(1952)] for the existence and the influence of this component is accompanied
by addition of physical arguments related to these defects. It is tried to
explain an underlying fluctuating spatial pattern of free volume with weak
contrast and a pattern of mobility with strong contrast, and to explain the
characteristic lengths for the main transition and the Fischer modes. The
structure of the relaxation chart (dynamic glass transition) and several
properties of, and relations between, the slower dispersion zones therein,
are reviewed for classical glassforming liquids of moderate complexity. For
the main transition, the preponderant component is pushed in the midst of
the defect and induces the molecule to its diffusion step across the cage
door of the next neighbors. An Experimentum Crucis for an indirect proof of
the existence of defects $-$ via characteristic lengths $-$ is also
described.
\end{abstract}

\maketitle

\newpage \newpage

\section{Introduction}

The Levy sum for stable distributions with Levy exponent $\alpha <1$ has a
preponderant component \cite{LevyBuecher,Darling1952,FellerBuch}. An example
of its physical relevance was recently published for Laser cooling \cite%
{BardouBuch}. The reviews for dynamics in classical molecular liquids are
sometimes accompanied by catalogs of open questions \cite%
{Ediger1996,AngellNgai2000,Angell2000,Ngai2000}. The relevance of Levy
distributions for this dynamics \cite{DonthBuch2001} may be discussed in a
thermodynamic frame of fluctuating space-time pattern for free volume, where
islands of mobility \cite{Johari1976} from temporary concentrations of free
volume can be formed by using a general length/mobility scaling \cite%
{DonthBuch1992}. Such spots in the pattern are called Levy defects, if they
can be explained by an appropriate Levy sum, in particular with a
preponderant component. A well known model is Glarum's defect diffusion
model \cite{Glarum1960} which was later reviewed \cite{Shlesinger1988} by
means of Levy flights, cf. also Ref.~\cite{Bouchaud1990}. The more later
variant of Bendler, Fontanella, and Shlesinger \cite{Bendler2003} is aimed
to interaction of several or many defects forming clusters as basis for the
glass transition. On the contrary, our variant \cite{DonthBuch2001} is aimed
to an explanation of molecular cooperativity of the dynamic glass transition
from one defect alone. An experimental indication for defects in larger
spatial scales are Fischer's speckles \cite%
{Fischer1993,Patkowski2001+Bakai2004}.\newline
\newline

If confirmed, the Levy defect pattern would be a surprising phenomenon: Upon
some ''primary democratic'' conditions, Levy statistics can promote shaping.%
\newline
\newline

The basic physical assumption of our approach is that the Levy sum
components can be related to spatially separable, dynamically independent
subsystems or partial systems. (Such a relation is not necessary, which
means that there are Levy situations where spatial aspects play a minor
role. An example is the treatment of Laser cooling in Ref.\cite{BardouBuch},
where a trap in the momentum space is considered, and the Levy distribution
is related to lifetimes such as trapping or escaping times.) Our assumption
is completed by a general space vs. mobility scaling of fluctuation modes to
treat the spatial structure of the defects. Our approach is supported by a
Representativeness Theorem, showing inversely that the distribution of
representative subsystems corresponds to a Levy sum. An Experimentum Crucis
for Levy defects of the main transition in classical liquids is suggested
that uses the relevant characteristic lengths.\newline
\newline

\begin{center}
$*\qquad *\qquad *$
\end{center}

As a rule, experimental retardation in glass forming liquids can well be
adjusted by a Kohlrausch function \cite%
{Kohlrausch1854,Williams1970,Goetze1992,NgaiJPCM2000} in the time $t$ domain,

\begin{equation}
\text{retardation }\sim \text{ correlation function }\sim \exp (-at^\alpha )%
\text{ , }a>0,\text{ }\alpha =0.4...1.0.  \label{Eq.1.1}
\end{equation}
The proportionality to a correlation function is a consequence of the
classical fluctuation dissipation theorem, FDT.\newline
\newline

On the other hand, such ''stretched exponentials'' are important in
probability theory: They are characteristic functions (Fourier transforms)
for stable (Levy) distributions with exponents $\alpha \leq 1$. If the
stretched exponentials Eq.(\ref{Eq.1.1}) are really indications for a Levy
distribution, then the latter is some kind of (inverse) Fourier transform of
such exponentials. Graphs for $\alpha \leq 1$ are in Ref.~\cite%
{DonthBuch2001}, pp.305-311.\newline
\newline

Since the Fourier transform of the above correlation function from the FDT
is a spectral density $x^2(\omega )$ for an extensive variable $x$
corresponding to the concrete retardation, the relevant Levy distribution
density is proportional to $x^2(\omega )~d\omega $,

\begin{equation}
f(\omega )~d\omega \sim x^2(\omega )~d\omega .  \label{Eq.1.2}
\end{equation}
This density is fractal at high frequencies,

\begin{eqnarray}
\text{spectral density }x^2(\omega ) &\sim &\omega ^{-1-\alpha },  \nonumber
\label{Eq.1.3} \\
\text{susceptibility ~~~~}\alpha ^{\prime \prime }(\omega ) &\sim &\omega
x^2(\omega )\sim \omega ^{-\alpha },  \label{Eq.1.3}
\end{eqnarray}
with the same exponent $\alpha $; $\alpha ^{\prime \prime }(\omega )$ is the
loss part of the susceptibility (dynamic compliance) corresponding to $x$, $%
\omega $ the frequency; $\log \omega $ is called the mobility. The general
concept of fractality is restricted here to a power law expressing
selfsimilarity.\newline
\newline

A Levy distribution with exponent $\alpha <1$ has neither a finite variance
{\bf D} nor a finite expectation {\bf E},

\begin{eqnarray}
{\bf D}(\omega ) &=&\text{Var}(\omega )\rightarrow \infty ~~\text{for~~}%
0<\alpha <2,  \nonumber \\
{\bf E}(\omega ) &\sim &\int\limits_{\omega >0}\omega ~x^2(\omega )~d\omega
\rightarrow \infty ~~\text{for~~}0<\alpha <1.  \label{Eq.1.4}
\end{eqnarray}
The second equation (\ref{Eq.1.4}) leads to a preponderant component in the
Levy sum for a Levy distribution considered as a stable limit distribution.
One term of the sum retains a finite influence of order the ''tilt'' ($%
1-\alpha $) on the sum, irrespective that the number of its components tends
to infinity.\newline
\newline

Are all these properties a direct consequence of the experimental Kohlrausch
function (\ref{Eq.1.1}), especially for short times? \cite{Fuchs1994}. Are
additional arguments necessary that cannot simply be concluded from the
existence of a Kohlrausch function or from present experiments at short
times $t$ or, correspondingly, at large frequencies $\omega $?\newline
\newline

As indicated above, we try here an answer via a defect pattern from free
volume $V^{\prime }(\omega )$, making higher mobility ($\log \omega $) at
places where more local free room is available for molecular motion. We
shall introduce a Levy sum via separated partial systems or thermodynamic
subsystems and try to explain the property $\alpha <1$ by some kind of
instability.\newline
\newline

It is a general difficulty in understanding liquid dynamics by an approach
that represents the complicated motion of molecules with ''strong
interaction'' between many of them by means of statistical independence,
being one of the basic assumptions to get a Levy distribution.\newline
\newline

The aim of this paper is a quasi thermodynamic approach to the dynamics of
classical liquids starting from a robust mathematical and physical
background for a fluctuating pattern with Levy defects. The mathematics used
is based on the classical theorems of Darling and Feller. The mathematical
development along their original proofs is accompanied by physical
arguments, because the handling of a probability distribution with
nonexistent expectation and variance (Eq.~(\ref{Eq.1.4})) is strange for
physicists which are educated and experienced by only distributions that
have normal (Gauss) limit distributions with finite, existing variances and
expectations, in particular with no preponderant components in the limit sum.%
\newline
\newline

\section{Terminology for our Levy Situation}

{\bf Aim}. A glossarial introduction of some verbal concepts seems useful
for the ''thermodynamic'' application of the Levy distribution in different
fields. Calling $f\,(x)\,dx=p(x)\,dx$ the density of a stable Levy
distribution, then examples for the $x$ coordinate of the sample space are: $%
x=\omega $ for the frequency of liquid dynamics, $x=k$ for the wave vector
of cosmic density fluctuation, and $x=$ money for the economics of a Levy
society (\cite{DonthBuch2001}, p.334-336). The discussion with the aid of
Levy distribution is more suitable for the macroscopic, phenomenological
thermodynamics (e.g. the state) and for susceptibilities near the
equilibrium (modulus $m(\omega )$, compliance $j(\omega )$); the discussion
with a Levy flight \cite{Bouchaud1990} is more suitable for a molecular
approach, e.g. the molecular diffusion leading to equilibrium values of a
chemical potential. Warning: This Section is not a summary of the difficult
probability theory of stable distributions \cite{FellerBuch}.\newline
\newline

{\bf Concepts}. (1) {\it Levy sum} \cite{FellerBuch} is defined by

\begin{equation}
{\bf S}_n={\bf X}_1+...+{\bf X}_n  \label{Eq.2.1}
\end{equation}
where the Levy sum components ${\bf X}_i$ are equivalent (cf. Point (2),
below) independent random variables. The sum is suitable for the discussion
of additive situations, e.g. for extensive thermodynamic variables,
compliances, or for money in economics. (For the index $i$ cf. Point (4),
below).\newline
\newline

(2) {\it Equivalence} of the Levy components means that their distributions $%
F_i(x)$ differ only by location ($\beta _i,b$) and scale or norming
parameters ($a_i,a$), e.g. for $i=1,2,$

\begin{equation}
F_2(x)=F_1(a_1x+\beta _1)\text{ , }a_1>0.  \label{Eq.2.2}
\end{equation}
Mathematically it is said that they are ''of the same type''. Another notion
for two distributions ${\bf U}$ and ${\bf V}$ is: ${\bf U}\stackrel{d}{=}a%
{\bf V}+b$. Often, mathematically, it is said that ''...there are constants $%
a_i$ and $\beta _i$ so that...''. In the applications we need robust reasons
for that: physical, cosmological, economical.... The ''same type'' is used
for general arguments, the ''parameters'' are used for individual
properties, e.g. for subsystems \newline
\newline

(3) {\it Exclusiveness. Levy scaling}. The former concept means that there
are mathematically stable limit distributions ($n\rightarrow \infty $) only
for Gauss ($\alpha =2$) and Levy ($0<\alpha <2$), with $\alpha $ the Levy
exponent (cf. Eq.~\ref{Eq.2.1a} and Point (4), below). For variations of
components (small $n$), there are domains of attraction to the limit only
for this two cases. It is important for the applications, that there is a
broad spectrum of possibilities which tend to only the two sharply defined
limits: Gauss or Levy, for the free volume or dynamics of liquids, or for
the invisible hand of economics. We have then, for the $n\rightarrow \infty $
limit distributions, only one behavior of norming constants. In the form of $%
{\bf S}_n\stackrel{d}{=}c_n{\bf X}_n+\gamma _n$, we get $c_n=n^{1/\alpha }$
(Levy scaling). A useful concept is the scaled Levy sum,

\begin{equation}
{\bf \tilde S}_n={\bf S}_n/n^{1/\alpha }.  \label{Eq.2.1a}
\end{equation}
This sum is existent for $n\rightarrow \infty $; the components ${\bf X}_i$
are then equivalent to ${\bf \tilde S}_n$. A norming factor $n^{-1/\alpha }$
is for ''damping'' the $n^{1/\alpha }$ ''increase'' of the original
nonscaled Levy sum ${\bf S}_n$.This damping is much stronger than for Gauss (%
$1/\sqrt{n}$ for $\alpha =2$). Examples are the sharp hierarchy of the
ordered Levy sum for $\alpha <1$, Ref.~\cite{BardouBuch}, or the
non-Arrhenius behavior of the dynamic glass transition at low temperature
(Section~VI.D), cf. also Theorem 1, below.\newline
\newline

(4) {\it Pluralism}. This concept describes the great variety of the Levy
distribution for applications. Two Examples. A. According to Eqs.~(\ref%
{Eq.1.4}), the Levy exponent $\alpha $ distinguishes three general cases
(apart from some special cases for $\alpha =2$ and $\alpha =1$): (a)
Expectation ${\bf E}$ and Variance ${\bf D}$ both finite: Gauss, $\alpha =2$%
; (b) ${\bf E}$ finite (''existent''), but ${\bf D}$ infinite
(''nonexistent'') for $1<\alpha <2$; and (c) both ${\bf E}$ and ${\bf D}$
infinite for $0<\alpha <1$; $1-\alpha >0$ is called a ''tilt''. The (b)
region is called hierarchy, the (c) region is called dictatorship (because
we find a preponderant Levy component there) on the top of a sharper
hierarchy. Example for (b): stock market with usually $\alpha \approx 1.5$.
Further. The Gauss case may be called democracy. The general equivalence of
the Levy sum components is called primary democracy. $-$ B. The index $i$ of
the components ${\bf X}_i$ in Eq.~(\ref{Eq.2.1}) must be completed by
further aspects: For the dynamics of liquids by temperature, pressure, kind
of response after different disturbances; for the Levy society by the field
(economics, politics, culture), and so on. The pluralism is a wide-ranging
concept: Even for the same situation or the same defect, different response
can have different exponents with different dependence on temperature,
pressure, and so on. Such ''individual'' $\alpha $ exponents are not fixed
by general arguments (cf. Section~VI.C, below).\newline
\newline

(5) {\it Fractality} characterizes here the power laws of Eq.~(\ref{Eq.1.3})
and their selfsimilarity. Defining \cite{CohenBuch1996} $F(x)$ or $F_x(x)$
as the distribution, $f(x)$ or $f_x(x)$ as the density with $%
F_x(x)=\int\limits_{-\infty }^xf_x(\xi )\,d\xi -$ $f(x)$ is continuous for
Levy distributions $-$ and $1-F_x(x)=\int\limits_x^\infty f_x(\xi )\,d\xi $
for large $x$ coordinates as tail, the fractility according to Eq.~(\ref%
{Eq.1.3}) holds for the large $x$, i.e. for the tail: $f(x)\sim x^{-1-\alpha
}$, with $\alpha $ the Levy exponent. Complementarily, for the
characteristic function (e.g. for the simpler Laplace transform instead of
Fourier, $\varphi (\lambda )$ or $\varphi _f(\lambda )=\int\limits_0^\infty
e^{-\lambda x}f(x)\,dx={\bf E\,}(e^{-\lambda {\bf X}})$), we get a stretched
exponential according to Eq.~(\ref{Eq.1.1}) with $t=$ time for Fourier and $%
x=\omega $ frequency in a spectral density $p(\omega )=f(\omega )$. The
stretched exponential is more suitable for large times (or for less money in
economy), where fractality is more suitable for high frequency (or many
money in economy). The input of a spatial aspect (Eq.~(\ref{Eq.3.4}), below)
leads to a spatial unity, called ''defect'', with, for $\alpha <1$, a
preponderant component in the midst, a fractal (hierarchical) center, and a
stretched, cooperative periphery, all with the same exponent $\alpha $. The
size of the defect may be defined by a cooperativity $N$.\newline
\newline

(6) {\it Levy instability}. Mathematically, the term ''stable'' means that,
whatever the deviations in a finite sum of independent random variables $%
{\bf X}_k$ in the domain of attraction are, the limit sum ${\bf S}_n$
stabilizes itself, i.e. ${\bf S}_n/n^{1/\alpha }$ converges to a definite
limit distribution for $n\rightarrow \infty $. Exclusiveness for $\alpha <2$
leads then to a Levy distribution. In the applications, for illustrative or
intuitive purposes, fractality for large $x$ coordinate values, esp. with a
preponderant component for $\alpha <1$, can be associated with a very
special kind of instability, called ''Levy instability'' \cite{DonthBuch2001}%
. Example for liquids: a local breakdown to high frequencies due to a center
of lower particle density, i.e. a local concentration of free volume. In the
applications, of course, the high frequencies ($\Omega $) are very large but
finite. ''Very large'' means a comparison of the center with the periphery,
and $\Omega \rightarrow \infty $ means a tendency, e.g. ${\bf D\,}(\omega
)\sim \Omega ^{2-\alpha }$ and ${\bf E\,}(\omega )\sim \Omega ^{1-\alpha }$,
formally with $\omega =\Omega \rightarrow \infty $. The liquid example is
described in Section~III.A.\newline
\newline

(7) {\it Levy situation} is defined as an applicative situation (e.g. in
physics, cosmology, economics) with a robust equivalence of the Levy
components in the Levy sum and with robust reasons for a Levy instability
for explanation of exponents $\alpha <2$, in particular of $\alpha <1$ when
a preponderant component is to be expected.\newline
\newline

{\bf Examples}. A distribution is symmetrical, if $f(x)=f(-x)$; its
expectation is zero, ${\bf E}({\bf X})=0$. The spectral densities for
classical (non-quantum mechanical) liquids are symmetrical, $x^2(-\omega
)=x^2(\omega )$. Positive variables ${\bf X}\geq 0$ (i.e. the variables $%
x\geq 0$) are often physically motivated. For our (and many general)
considerations about Levy distribution, ${\bf X}\geq 0$ is no serious
restriction, because centering (selection of $\gamma _n$ in Point (3) above)
is relatively free in the frame of equivalence. [To find e.g. the general
expression for Levy distributions, depending on $\alpha $ and $\gamma $
(Ref.~\cite{FellerBuch}, p.580), no centering procedure is required for $%
\alpha <1$, while for $\alpha >1$ the natural centering to zero expectation
suffices. In Particular, the influence of preponderant component is
independent of $\gamma $ values. For construction of a concrete moleculare
structure of a defect, however, centering may be important~\cite%
{Bouchaud1990}.]\newline
\newline

For the sum of two independent variables, ${\bf X}_1+{\bf X}_2$, the joint
probability density $f_g~(x_1,x_2)=f_1(x_1)\cdot f_2(x_2)$. A common
variable may be defined by the sum $x=x_1+x_2$. The common distribution $%
F(x) $ for $P(\xi <x)$ is obtained by integration over the $x$ space region $%
x<x_1+x_2$ with measure $dx_1dx_2$. We get then for the sum density the
convolution

\begin{equation}
f=f_1*f_2,\quad \text{i.e.}\quad f(x)=\int f_1(z)~f_2(x-z)~dz.
\label{Eq.2.3}
\end{equation}
\newline
\newline

For the ratio of independent variables, ${\bf X}_1/{\bf X}_2$, we have e.g.
a joint probability density $f_g(x_1,x_2)\sim x_2~f_1(x_1x_2)~f_2(x_2)$ for $%
F(x)=P\{\frac{x_1}{x_2}<x\}$ \cite{GnedenkoBuch}. The integration is more
complicated than for the sum, because the ratio of the variables $x_1/x_2<x$
must be used. The result is

\begin{equation}
f(x)=\int\limits_{-\infty }^\infty \left| z\right| ~f_1(zx)~f_2(z)~dz
\label{Eq.2.4}
\end{equation}
with the possibility of a partial rescaling procedure that affects $f_1$ and
$f_2$ differently: the product $x_1x_2$ vs. $x_2$ alone.\newline
\newline

Be $f_1$ a symmetrical Gauss density ($\alpha =2$) with variance ${\bf D}%
_1=\sigma _1^2$, and $f_2$ such a density with ${\bf D}_2=\sigma _2^2$,
then, for independent variables, the sum gives again a symmetrical Gauss\
density with variance ${\bf D}=\sigma ^2=\sigma _1^2+\sigma _2^2$ (Gauss\ +
Gauss = Gauss). Their ratio, however, gives a Cauchy distribution,

\begin{equation}
f(x)=\frac{\sigma _1\sigma _2}{\pi (\sigma _1^2+\sigma _2^2x^2)}.
\label{Eq.2.5}
\end{equation}
This is a symmetrical Levy distribution with exponent $\alpha =1$, i.e. with
no existing variance (${\bf D}=\infty $), and applied for positive variables
$x\geq 0$, also with no expectation (${\bf E}=\infty $); (Gauss/Gauss =
Cauchy). The ratio favours large variances, but a Gauss/Gauss ratio cannot
generate preponderant components. The latter are restricted to Levy
exponents $\alpha <1$.\newline
\newline

\section{Fluctuating pattern with Levy defects}

This section is to describe pictures for Levy instability and for
equivalence of Levy sum components, i.e. a Levy situation for the
fluctuating pattern of molecular liquids.

\subsection{Free volume. Partial systems. Levy instability}

To get a Levy instability, we consider a local breakthrough of molecular
mobility ($\log \omega $) as response to a virtual local concentration of
free room for molecular movements. The aim of the concept of free volume ($%
V^{\prime }$) for the pattern is that a local concentration of free volume,
i.e. a low local particle density, pushes the mobility there to high
frequencies, in the limit to $\omega $ of order $\Omega \approx 10^{12}$%
rad/s as for free rattling of particles in a sufficiently large molecular
cage of next neighbors. Small local density reduction of order some percent
is sufficient to rise $\omega $ from e.g. 10$^6$rad/s in the main transition
to cage rattling, because the repulsive inter (and relevant intra) molecular
potentials are steep. Locality of the breakthrough follows from a given free
volume of the total sample in equilibrium. Lower local density here demands
higher local density somewhere else. Lower density everywhere would finally
lead to negative pressure. As the degree of divergence in expectation ${\bf E%
}$ and variance ${\bf D}$ is larger for smaller Levy exponents $\alpha $, we
may say that this exponent controls the Levy instability.\newline
\newline

To get a large number of equivalent small systems as components of a Levy
sum we partition the volume of a subsystem into small parts, called {\it %
partial systems}: The large numbers needed for a robust dynamic partition
for the limit distribution of spectral density $x^2(\omega )$ are thought to
be numbers of random attempts for rearranging the molecular situation in the
nanometer range there. We have many high-frequency attempts for a relevant
low-frequency ($\omega $) rearranging in the dispersion zone considered; the
number of attempts (as used for the limit) in the time interval 1/$\omega $
is much larger than the number of affected particles. We think about
molecular collisions which may cause the emission of a quantum $\hbar \omega
$ that can be catched by Nyquist's \cite{Nyquist1928} transmission lines in
a model for thermodynamic response, cf. also Ref.~\cite{DonthBuch2001},
p.~270.\newline
\newline

Be careful to distinguish the two possibilities. The large number $n$ for
the spatial systems leads to a Gauss distribution for free volume with weak
contrast in the pattern, the large number $n$ for the attempts leads to a
Levy distribution with exponent $\alpha <1$ for the spectral density with
strong contrast. The relation between the two is explained in Step 2 of the
Proof for the Representativeness Theorem in Section~VI.D, below.\newline
\newline

We now assume that the frequency of slow rearranging ($\omega $) is locally
controlled by the local free volume ($V^{\prime }$). For given independent
partial systems (index $i$) we assume, therefore, the existence of functions
$\omega _i(V_i)$ for each partial system,

\begin{equation}
\omega _i=\omega _i(V_i^{\prime })  \label{Eq.3.1}
\end{equation}
(no sum convention; Fig.~1). This assumption is justified by the high
attempt rate. If the partial system is not too small, i.e. if the attempt
rate inside is still much larger than the rearranging frequency,

\begin{equation}
\omega \ll \text{ attempt rate }\lesssim \Omega \approx 10^{12}\text{rad/s,}
\label{Eq.3.2}
\end{equation}
then we find a limit distribution for each partial system. Robustness of
equivalence and partition for the domain of attraction and exclusiveness of
Levy distribution allows to find a local control (\ref{Eq.3.1}) and to
identify the partial systems \{$i,i=1,...,n,$\} with independent components
of a Levy sum for free volume,

\begin{equation}
{\bf S}_n^{\prime }=V_1^{\prime }(\omega _1)+...+V_n^{\prime }(\omega _n).
\label{Eq.3.3}
\end{equation}

\begin{figure}[here]
\begin{center}
\includegraphics[width=0.5\linewidth]{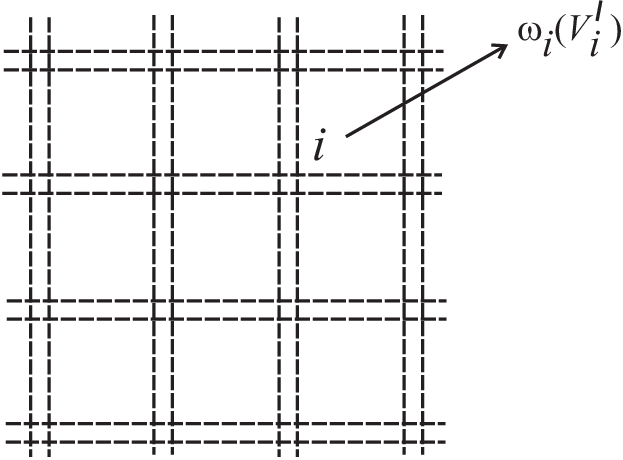}
\caption{Local control of mobility ($\log \omega $) by free volume ($%
V^{\prime }$) of partial systems, or by free volume of
representative subsystems (index $i$, Eq. (\ref{Eq.3.1})).
\label{fig1}}
\end{center}
\end{figure}

\subsection{Cooperativity}

A problem for intuition is whether the statistical independence of partial
systems with infinitive expectation (\ref{Eq.1.4}) due to high attempt rates
is sufficient for explanation of a phenomenon that is usually described by
cooperativity or cooperative rearrangement \cite{AdamGibbs1965}. In the
limit, each component has an equivalent dynamic distribution $x^2(\omega )$
[that is of the same type as all others, including the sum]. This is valid
also for partial systems in the periphery (\ref{Eq.3.4}) of the defect.
Their contribution to the sum (general susceptibility (\ref{Eq.1.2})) may be
small, nevertheless the movements in these partial systems may become very
lively because of the infinite expectation (\ref{Eq.1.4}), ${\bf E}%
\rightarrow \infty $. Any partial system can profit from the possibilities.
It can e.g. take the advantage to use the high frequencies for an effect on
the rearrangement, e.g. to become preponderant and get the center: All
particles may take part in diffusion.\newline
\newline

Competition for cooperativity can so be simulated by independent attempts of
partial systems. Expressed e.g. by the Gauss distribution for free volume,
large fluctuation of one partial system $i$ must be compensated by small
fluctuation of some others, because the fluctuation of the sum becomes
relatively small ($1/\sqrt{n}$). We get the chance to determine the extent
of cooperativity ($N_{mt}$) from thermodynamic variables (Section~VII.A,
below). For the Levy distribution with exponent $\alpha <1$ for dynamics,
the competition of the interchangeable, equivalent Levy sum components leads
to a preponderant component, probably connected with the diffusion step
through the cage door, necessary for equilibrium in liquids.\newline
\newline

How can cooperativity be distinguished from collectivity, sometimes also
used \cite{Goetze1992} for the glass transition? Two types of collectivity
are used for glass transition: (I) A common treatment of the high frequency $%
c$ and $a$ processes \cite{Goetze1992,Goetze1984} (Fig.~3, below), and (II)
A certain structure of thermodynamic phases (order parameter, clustering %
\cite{Bendler2003,Patkowski2001+Bakai2004}).\newline
\newline

Let us here discuss type (I) collectivity. The separation of $c$ and $a$
process is related to two neighbored processes that both are at high
frequencies and have direct mechanical relations. A G\"otze ansatz can,
therefore, use one formula for both, e.g. $m(t)=v_1\phi (t)+v_2\phi ^2(t)$,
where $m(t)$ is a memory and $\phi (t)$ is the relevant correlator, the mode
of the mode coupling theory MCT. This leads to a cusp bifurcation into $a$
and $c$ and to several scaling properties around the bifurcation, with
reasonable experimental confirmation. Collectivity is therefore a suitable
concept for the $a$ and $c$ processes in the relaxation chart of dynamic
glass transition.\newline
\newline

What, however, about the crossover region from the $a$ to the $\beta $ and $%
\alpha $ processes (Fig.~3, below)? This is no bifurcation (Section VI.E
below), which was assumed by MCT at the beginning. The rare experimental
information about the crossover \cite{Beiner1998,Kahle1999} shows two
scenarios different from a bifurcation. It is discussed below (Section VI.E)
that the defect pattern alone suffices to explain the existence and location
of the Johari Goldstein $\beta $ process. We see that collectivity is not
needed for understanding the $\alpha ,\beta $, and $\phi $ processes of the
dynamic glass transition.\newline
\newline

Remark. The above ''inside'' argumentation via partial systems for obtaining
a Levy distribution is consistent to an ''outside'' proof via minimal
subsystems (Fig.~4, Representativeness theorem Section VI.D).\newline
\newline

\subsection{Johari's islands of mobility}

The concept of islands of mobilities is introduced by Johari \cite%
{Johari1976}. In our context, we argue as follows.\newline
\newline

Since the free volume ($V^{\prime }$) has finite variance and expectation,
there is no preponderant component in its distribution: we find a Gauss
distribution for $V^{\prime }$. For a dynamic spectral density $x^2(\omega )$
of the ''same'' partition in partial systems we obtain, however, a
preponderant component due to the Levy instability (Step 2 of Section VI.C).
A spatial concentration of free volume is, however, possible, if the
interchangeable free partial or sub systems can spatially be redistributed.
Some shallow ''defect'' in a free volume pattern (as basis for a mobility or
frequency pattern from Eq.~(\ref{Eq.3.1})) can then be constructed by means
of a general length/time or length/mobility scaling \cite{DonthBuch1992} for
relevant modes (Fig.~2a and 2b),

\begin{figure}[here]
\begin{center}
\includegraphics[width=0.5\linewidth]{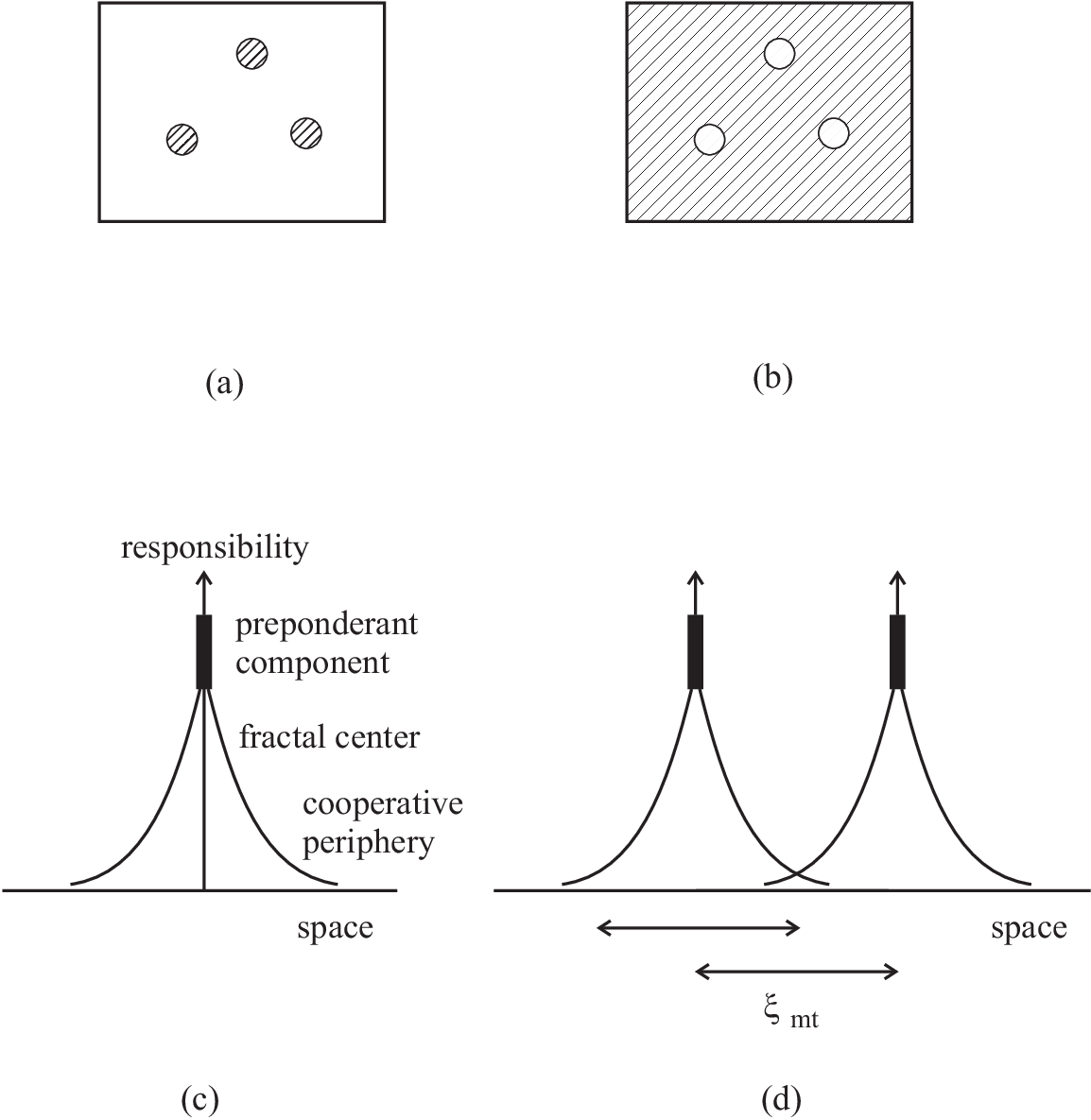}
\caption{(a + b). Defect pattern. If the white regions have more
free volume, and the hatched regions less free volume, then (a) =
islands of
immobility that are not consistent with the general length/time scaling (\ref%
{Eq.3.4}), but (b) = islands of mobility are consistent
\cite{Donth1996} (dynamic heterogeneity). (c). Sketch of the
spatial defect structure. Calling the spatially separable
contributions to the susceptibility $\alpha ^{\prime \prime
}(\omega )\sim \omega x^2(\omega )$ the ''responsibility'' [with
$x^2(\omega )$ the spectral density via the FDT], we obtain from
the above scaling and the Levy distribution a sharpened spatial
picture for one defect. The one Levy distribution corresponds to
the three parts of the defect: the influence of the preponderant
component (for $\alpha <1$), the fractality of the center, and the
cooperativity of the periphery. (d). The size of the defect in the
pattern is determined by the average distance (Section VI.B): The
statistical independence of minimal representative
subsystems for the $mt$ dispersion zone determines a characteristic length $%
\xi _{mt}$ via von Laue thermodynamics (Section VII.A).
\label{fig2}}
\end{center}
\end{figure}

\begin{equation}
\left.
\begin{array}{c}
\text{defect periphery : large mode length}\sim \text{ low mobility} \\
\text{defect center : small mode length }\sim \text{ large mobility}%
\end{array}
\right\}  \label{Eq.3.4}
\end{equation}
This means that the $\omega (V^{\prime })$ function of (\ref{Eq.3.1})
increases very sharply at the upper boundary for local free volume
fluctuations. We get the picture of Fig.~2c. In the center of the defect we
find the partial systems with much free volume, in the periphery those with
less free volume. Transformation to the frequency (\ref{Eq.3.1}) means that
the preponderant component related to the spectral density is near or in the
midst of the center for this island of mobility. In summary, we find more
free volume in the center of the defect.\newline
\newline

Let us underline again, that the density differences across the defect are
small, since the repulsive intermolecular potentials are steep. We get a
{\it dynamic heterogeneity} with large mobility differences (but no or a
small ''structural'' heterogeneity): a fluctuating pattern with temporary
mobility defects. In terms of point (4) of Section~II, the Johari Goldstein $%
\beta $ process (Section~VI.E) contributes to the dynamic dictatorship of
the preponderant component.\newline
\newline

\subsection{Description of the defect}

{\bf 1.} The defect embraces three properties of the Levy distribution
(Fig.~2.c): (i) Fractality properties (\ref{Eq.1.3})-(\ref{Eq.1.4}) that
care for the influence of large frequencies in the spectral density. (ii). A
stretched exponential (\ref{Eq.1.1}) for the correlation function. (iii).
The preponderant component of the Levy sum for spectral density.\newline
\newline

Assuming the above space/time scaling (\ref{Eq.3.4}), then the Levy
distribution stabilizes the Levy instability by connecting the center of the
defect (fractal divergencies at high frequencies, preponderant component in
the midst of the center, collection of partial systems with more free
volume) with its cooperative periphery (long-time tail of relaxation from
the stretched exponential, collection of partial systems with less free
volume). Embracing by one distribution with one exponent $\alpha $ results
in the correspondence between center and periphery (without any
freezing-in). The statistical independence and equivalence of the Levy sum
components inside the defect describes spatial aspects of the competition
for cooperativity, e.g. for being the preponderant component in the midst of
the center or for belonging to the hierarchy around the midst.\newline
\newline

{\bf 2.} We have three aspects of the preponderant component in the defect.
(a) Induction of an extraordinary process for volume and entropy
fluctuation. (b) The $n\rightarrow \infty $ limiting process for the Levy
sum pushes the preponderant component in the midst of the center (Sections
V.B-C). (c). The molecular picture is the diffusion step through the cage
door and the promotion of the Johari Goldstein process.\newline
\newline

{\bf 3.} The size of the defect results from the length/time scaling (\ref%
{Eq.3.4}). We find an opposite behavior of responsibilities (Fig. 2c) on the
way from a given defect to a neighbor defect. The responsibility in the
periphery of the first defect decreases and becomes smaller than the
increasing responsibility in the periphery of the second defect (Fig.2d). We
get an average distance in the pattern. The absolute length (in nanometers)
can be determined by thermodynamics, if there are reasons to consider
equivalent defects as some minimal representative units (Sections VI.C and
VII.A, below).\newline
\newline

The three points 1.-3. for the defects allow some visualization for the
shaping power of Levy statistics in form of a fluctuating mobility or
responsibility pattern in liquids.\newline
\newline

\section{Relationship between Kohlrausch function, Levy sum, and Levy
instability. Theorem 1}

A Kohlrausch correlation function (\ref{Eq.1.1}) can also, in principle, be
imagined without a relationship to Levy distributions, e.g. as obtained from
a dynamic differential equation or even from a Hamiltonian. As mentioned in
Section II, additional things are necessary to relate Eq.~(\ref{Eq.1.1})
with a Levy distribution for spectral density ($x=\omega $). This is
expressed by the ''if'' in the following (Ref.~\cite{FellerBuch}, p.~448).%
\newline
\newline

{\bf Theorem 1}. For fixed Levy exponent $\alpha $, $0<\alpha <1$, the
function $\varphi ^{(\alpha )}(\lambda )=\exp (-\lambda ^\alpha )$ is the
Laplace transform of a distribution $F^{(\alpha )}(x)$ with the following
properties:

(1) $F^{(\alpha )}(x)$ is a Levy distribution; more precisely, if ${\bf X}%
_1,...,{\bf X}_n$ are independent random variables with distribution $%
F^\alpha (x)$, then the normalized (scaled) Levy sum $({\bf X}_1+...+{\bf X}%
_n)/n^{1/\alpha }$ has again the distribution $F^{(\alpha )}(x)$.

(2) Fractality is obtained that may be normed by\qquad
\begin{equation}
x^\alpha [1-F^{(\alpha )}(x)]\rightarrow 1/\Gamma (1-\alpha ),x\rightarrow
\infty .  \label{Eq.4.1}
\end{equation}

{\bf Comment}. According to our basic assumption (Section I) we think here
about spatially separable partial systems when components of a Levy sum are
considered. For a physical Levy situation we need, for Theorem 1, a robust
partition in equivalent partial systems, and, for Levy exponent $\alpha <1$,
a Levy instability. In general, more trivially, it is the sum that allows
exclusiveness of the Levy distribution to become a limit distribution for
large numbers $n$.\newline
\newline

{\bf Proof} of Theorem 1. Let us first of all recall that the probability
density for a sum of independent variables may be transformed to a
convolution (\ref{Eq.2.3}) of the component densities that are (Fourier or)
Laplace transformed in a product. Therefore, in the limit we get for any
stable distribution an infinitely divisible distribution. The scaling
property $n^{1/\alpha }$ of stability (1) follows from infinite divisibility
of the Kohlrausch function (stretched exponential (\ref{Eq.1.1})): $(\varphi
^{(\alpha )})^n(\lambda )=\varphi ^{(\alpha )}(n^{1/\alpha }\lambda )$.$-$
The function $\varphi ^{(\alpha )}$ is completely monotone. Since $\varphi
^{(\alpha )}(0)=1$, the measure $F^{(\alpha )}(x)$ with Laplace transform $%
\varphi ^{(\alpha )}$ has a total mass 1.\newline
\newline

The proof for part (2) is obtained along the following line. A positive
function $L$ defined for $0<x<\infty $ {\it varies slowly at} $\infty $, if
for every fixed $x$

\begin{equation}
\frac{L(ax)}{L(a)}\rightarrow 1~\text{ for ~}a\rightarrow \infty .
\label{Eq.4.2}
\end{equation}
Such functions may be used to describe the domain of attraction of a
probability distribution. The limit of Eq.~(\ref{Eq.4.2}) is only fulfilled
by power functions (the basis for our fractility) which is, after a special
selection of the probability by the Gamma function of the tilt ($1-\alpha $%
), $\Gamma (1-\alpha )$, expressed by (\ref{Eq.4.1}). End of the Proof.%
\newline
\newline

\section{Physical understanding of the preponderant component for Levy
defects}

Let us recall Feller's (Ref.~\cite{FellerBuch}, p. 172) mathematical
heuristic. ''Consider, for example, a stable distribution ... with $\alpha
<1 $.... The {\it average} $({\bf X}_1+...+{\bf X}_n)/n$ has the same
distribution as ${\bf X}_1n^{-1+1/\alpha }$, and the last factor tends to $%
\infty $. Roughly speaking we can say that the average of $n$ variables is
likely to be considerably larger than any given component ${\bf X}_k$. This
is possible only if the {\it maximal term} ${\bf M}_n=\max [{\bf X}_1,...,%
{\bf X}_n]$ is likely to grow exceedingly large and to receive a
preponderating influence on the sum ${\bf S}_n$''. Section V is firstly to
bring some analysis used and is secondly to describe the preponderant
component in Levy defects more precisely than in Section III.\newline
\newline

\subsection{Darling Lemma}

To calculate the influence of a preponderant component as expressed by the
Levy exponent $\alpha <1$ we need a method for handling the maximal
component (${\bf M}_n$) in a Levy sum (${\bf S}_n$). This is demonstrated
along the original Proof for following (Darling's \cite{Darling1952})\newline

{\bf Lemma}. The Laplace transform $\varphi _z(\lambda )\equiv \xi
_n(\lambda )$ of the random variable ${\bf Z}_n={\bf S}_n/{\bf M}_n$ (which
ratio should characterize the above influence) is

\begin{equation}
\xi _n(\lambda )=n\,e^{-\lambda }\int\limits_0^\infty ~(\beta
\int\limits_0^1e^{-\lambda \gamma }~f~(\gamma \beta )~d\gamma
)^{n-1}~f\,(\beta )~d\beta  \label{Eq.5.1}
\end{equation}
with ($\beta ,\gamma $) dummy variables for the probability density $f$ of
equivalent components \{$X_i$\}.\newline
\newline

{\bf Proof}. We start from the equivalence of the Levy components. There is
no loss in generality by assuming ${\bf X}_1={\bf M}_n$, since each ${\bf X}%
_i$ has a probability of $1/n$ of being the largest term (and $%
P\{x_i=x_j\}=0 $ for $i\neq j$ since the distribution is presumed to be
continuous). The joint density $f_g$ (index $g$) is then

\begin{equation}
f_g\,(\beta _1,\beta _2,...,\beta _n)=\left\{
\begin{array}{c}
n\,f\,(\beta _1)~f\,(\beta _2)...f\,(\beta _n) \\
\qquad \qquad \qquad \text{ if }\beta _1=\max \{\beta _i\} \\
0\text{\qquad otherwise}%
\end{array}
\right.  \label{Eq.5.2}
\end{equation}
Then the relevant expectation $\xi _n=$

\begin{eqnarray}
{\bf E\,}(e^{-\lambda {\bf Z}_n}) &=&\int \int\limits_0^\infty ...\int
e^{-\lambda (x_1+x_2+...+x_n)/x_1}\times  \nonumber \\
&&\times f_g(x_1,x_2,...,x_n)~dx_1\,dx_2...dx_n  \nonumber \\
&=&ne^{-\lambda }\int\limits_0^\infty \int\limits_0^\beta
...\int\limits_0^\beta e^{-\lambda (\gamma _2+...+\gamma _n)/\beta }\times
\nonumber \\
&&\times f\,(\gamma _2)...f\,(\gamma _n)\cdot f\,(\beta )\cdot d\gamma
_2...d\gamma _nd\beta  \nonumber \\
&=&ne^{-\lambda }\int\limits_0^\infty ~\left\{ ~\int\limits_0^\infty
e^{-\lambda \gamma /\beta }f(\gamma )\,d\gamma \right\} ^{n-1}f(\beta
)\,d\beta .  \label{Eq.5.3}
\end{eqnarray}
The Lemma follows from rescaling of the dummy variables: $\beta d\gamma
\rightarrow d(\gamma \beta )$. The upper limit 1 in the inner integral comes
from $\gamma \leq \beta $, i.e. from the maximality of the component for $%
\beta $. Note the different variables, $\beta $ and $\gamma \beta $, for the
densities $f$ in the Lemma (\ref{Eq.5.1}), typical for ratios of random
variables (\ref{Eq.2.4}).\newline

\subsection{Influence of the preponderant component. Theorem 2}

This subsection is to calculate the expectation of the random ratio: Levy
sum over its maximal component, ${\bf Z}_n={\bf S}_n/{\bf M}_n$ with
variable (observable) $\xi _n$, for $n\rightarrow \infty $.

{\bf Theorem 2} (Darling \cite{Darling1952}). The expectation of ${\bf Z}_n$
for the limit $n\rightarrow \infty $ with a Levy exponent $0<\alpha <1$ is

\begin{equation}
{\bf E}\,({\bf S}_n/{\bf M}_n)\rightarrow 1/(1-\alpha ).  \label{Eq.5.4}
\end{equation}
The {\bf Proof} is divided in two parts: The limit is first explicitly
calculated for the limit distribution itself, and in a Comment the
participation of the domain of attraction is discussed. We put for the inner
integral of the Laplace transform $\varphi _z^{(n)}(\lambda )=\xi _n(\lambda
)={\bf E}(e^{-\lambda {\bf Z}_n})$ from the Lemma, Eq.~(\ref{Eq.5.1}),

\begin{equation}
\psi (\beta )\stackrel{\text{def}}{=}\beta \int\limits_0^1e^{-\lambda \gamma
}f\,(\gamma \beta )~d\gamma .  \label{Eq.5.5}
\end{equation}
Note the different variables in exponent and density. In the limit ($%
n\rightarrow \infty $), only the tails of the Levy distribution are
important (cf. the Comment). Hence we put for their fractility

\begin{equation}
1-F(x)=\tilde c/x^\alpha  \label{Eq.5.6}
\end{equation}
with a constant $\tilde c=\tilde c(\alpha ,\tilde \gamma )$ where $\tilde %
\gamma $ is ascribed to centering. There are no further parameters beyond ($%
\alpha ,\tilde \gamma $) in the general Levy density. Eq.~(\ref{Eq.5.6}) is
a consequence of Eqs.~(\ref{Eq.4.1}) and (\ref{Eq.4.2}). Then, from
arithmetic manipulation in direction of the tails, such as

\begin{equation}
(1-F(\gamma \beta ))-(1-F(\beta )),  \label{Eq.5.7}
\end{equation}
we get

\begin{equation}
\psi (\beta )=1-(1-F(\beta ))-\lambda \int\limits_0^1e^{-\lambda \gamma
}\,[(1-F(\gamma \beta ))-(1-F(\beta ))]~d\gamma .  \label{Eq.5.8}
\end{equation}
The integral yields

\begin{equation}
\text{integral }\stackrel{\text{def}}{=}-\lambda \int\limits_0^1e^{-\lambda
\gamma }\,[...]~d\gamma =\frac{\tilde c}\beta \,\alpha
\int\limits_0^1(e^{-\lambda \gamma }-1)\,\frac{d\gamma }{\gamma ^{\alpha -1}}%
.  \label{Eq.5.9}
\end{equation}
Defining the number

\begin{equation}
\phi _1\stackrel{\text{def}}{=}\alpha \int\limits_0^1(e^{-\lambda \gamma
}-1)\,\frac{d\gamma }{\gamma ^{\alpha -1}}  \label{Eq.5.10}
\end{equation}
we obtain

\begin{equation}
\psi (\beta )=1-(1-\phi _1)\,\tilde c/\beta ^\alpha .  \label{Eq.5.11}
\end{equation}
Rescaling now with the variable

\begin{equation}
v\stackrel{\text{def}}{=}\tilde c/\beta ^\alpha ,  \label{Eq.5.12}
\end{equation}
(which eliminates $\tilde c$ and therefore any centering $\tilde \gamma $ in
$\tilde c$), gives in our limit

\begin{equation}
\xi (\lambda )=\frac{e^{-\lambda }}{1-\alpha \int\limits_0^1(e^{-\lambda
\gamma }-1)\,\gamma ^{-1-\alpha }\,d\gamma }.  \label{Eq.5.13}
\end{equation}
The wanted expectation value follows as

\begin{eqnarray}
{\bf E\,}(\xi (\lambda )) &=&-\xi ^{\prime }(0)=-d\xi \,/\,d\lambda
\,(\lambda =0)=  \nonumber \\
&=&-\lim\limits_{\lambda \rightarrow 0}~\frac d{d\lambda }\,(1-\lambda
+\lambda \frac \alpha {\alpha -1})=\frac 1{1-\alpha }.  \label{Eq.5.14}
\end{eqnarray}
The influence of the maximal term on the limit Levy sum is calculated as $%
1/(1-\alpha )$ in the ratio ${\bf S}_n/{\bf M}_n$. The only remaining
parameter is the Levy exponent $\alpha $.\newline
\newline

We may say, that the influence of the maximal term, the preponderant
component, is of order ($1-\alpha $), (the tilt in cosmological terms).
Compared with Gauss, where the influence of the maximal term remains
infinitely small, we get e.g. for $\alpha =1/2$ the surprising result, that
the maximal, the preponderant component has approximately the same influence
as the rest of the infinitely many components.\newline
\newline

{\bf Comment}. The domain of attraction is handled by the deviation from
Eq.~(\ref{Eq.4.2}), using the symbol $o(1)$ which means of the order less
than 1,

\begin{equation}
\frac{L(ax)}{L(a)}=1+o(1).  \label{Eq.5.15}
\end{equation}
We pursue this symbol along the proof by the chain due to rescalings,

\begin{equation}
o(1)\rightarrow o\,(1-F(\beta ))\rightarrow o(1/n).  \label{Eq.5.16}
\end{equation}
From (\ref{Eq.5.5}) we get for $1-F(\gamma \beta )$ after $\gamma \beta
\rightarrow \gamma $ rescaling

\begin{equation}
1-F(\gamma \beta )=(1/\gamma ^\alpha )\,(1-F(\gamma )(1+o(1)).
\label{Eq.5.17}
\end{equation}
The integral (\ref{Eq.5.9}) yields then the

\begin{equation}
\text{integral }=(1-F(\beta ))~\alpha \int\limits_0^1(e^{-\lambda \gamma
}-1)~\gamma ^{-1-\alpha }\,d\gamma +o\,(1-F(\beta )).  \label{Eq.5.18}
\end{equation}
The shift into the tail is therefore

\begin{equation}
\psi (\beta )=1-(1-\phi _1)\,(1-F(\beta ))+o\,(1-F(\beta ))  \label{Eq.5.19}
\end{equation}
with the same $\phi _1$ as before, Eq.~(\ref{Eq.5.10}). Since $\left| \psi
(\beta )\right| <1$ for bounded $\beta $, the early portion of the integral
for $\xi _n(\lambda )$ is negligible for sufficiently large $n$. This can be
seen after introduction of a new variable instead of (\ref{Eq.5.12}), now

\begin{equation}
v=n(1-F(\beta )).  \label{Eq.5.20}
\end{equation}
Hence

\begin{equation}
\xi _n(\lambda )\sim e^{-\lambda }\int\limits_0^n(1-\frac vn\,(1-\phi
_1)+v\,o\,(\frac 1n))^{n-1}\,dv,  \label{Eq.5.21}
\end{equation}
and we get again $1/(1-\alpha )$ in the limit. End of the Proof for
Theorem~2.\newline
\newline

We see from the Comment chain (\ref{Eq.5.16}) that the preponderant
component really pushes the play into the tail's fractality asymptote. The
large number of participants (attempts in our model) pushes the preponderant
component into the midst of the defect center; Fig.~2c is a picture for the
shaping power of the Levy statistics.\newline
\newline

Remark. Absence of a preponderant component would mean ${\bf S}_n/{\bf M}%
_n\rightarrow \infty $. This is prevented by the $o(1/n)$ pushing into the
defect center and the finiteness of the remaining integral due to the
fractality $\gamma ^{-1-\alpha }$ in the denominator of (\ref{Eq.5.18}).
This leads to the finiteness of $-\xi ^{\prime }(o)$ in (\ref{Eq.5.14}).%
\newline

\subsection{Verbal understanding of a preponderant component in a defect}

This brings some comments to the Points (4) and (6) of Section~II and
continues the defect discussion of Section III. The Levy distribution is
characterized by a large number ($n$) of statistically independent
equivalent random components in a sum (inside example: the partial systems
of a subsystem; outside example: the subsystems of a larger subsystem) and a
Levy exponent $\alpha $ that describes some Levy instability, larger for
smaller $\alpha $. Let us remark, that the equivalence of the Levy sum
components means that they are interchangeable near equilibrium. They can be
interchanged between the periphery, the hierarchy, and the preponderance of
the center.\newline
\newline

In a spatial picture of a $\alpha <1$ defect (Fig.~2c above), the
preponderant component is in the midst of the fractal center of the defect.
This component is pushed in the center by Eq.~(\ref{Eq.5.16}). The defect
center collects the components with more influence around the preponderant
component: the sharp ''hierarchy'' below the ''dictator''. The other
components are in the cooperative periphery where the components with
temporarily less influence on the sum are collected, although all components
are equivalent by having infinite expectation ($E\rightarrow \infty $). The
Levy distribution stabilizes the defect by connecting the fast fractal
center with the slow cooperative periphery; there is a correspondence
between center and periphery. The cooperativity is stimulated (encouraged)
by the equivalence in expectation. For infinite expectation, the competition
of the equivalent and independent components is stimulated by the property
that a considerable part of influence, of order the tilt ($1-\alpha $), on
the total sum is realized by only one, the preponderant component,
interchangeable with the others in the equilibrium.\newline
\newline

In other words, the ''actual events'' are pushed in the center with the
preponderant component. This are therefore fast actions. In liquids, the
preponderant component in dynamics may induce the diffusion step through the
molecular cage door as basic process to maintain the equilibrium. In the
midst of the center there is only action for one, the preponderant
component. [The ''dictator'' can fill the midst because of his infinite
''personal'' expectation, $E\rightarrow \infty $.] This is mathematically
complemented by the absence of (this) one component outside, which absence
makes the integrals of Eq.~(\ref{Eq.5.1}) suitable.\newline
\newline

Let us conclude with an example for nonequilibrium. If we have an underlying
trend, in liquids e.g. a cooling rate $\dot T=dT/dt<0$, then we may ask,
which parts of the defect firstly cannot longer follow. Due to the general
length/time scaling (\ref{Eq.3.4}) relevant for the spatial structure of the
defect, it is the slow periphery that first falls out of equilibrium. The
trend leads unavoidably to incrustation, to a frozen vault structure (Ref.~%
\cite{DonthBuch2001}, p. 386), while the center continues a time with
responsibility to outer dynamic contacts.\newline
\newline

\section{Dynamic glass transition in moderate molecular liquids}

\subsection{Dispersion zones in a mobility -(reciprocal) temperature diagram}

The term ''dynamic glass transition'' is synonymously used for the dynamics
in liquids. For classical molecular liquids of moderate complexity
(''moderate liquids'') we find a surprisingly general relaxation chart, i.e.
a $\log \omega -1/T$ diagram (Fig.~3). There is a characteristic arrangement
of dispersion zones (zones of higher dissipation) with a typical width of
one or a few frequency decades, if crystallization can be prevented. The
symbols and names of these zones are listed in the figure caption. This
picture can be discussed as a consequence \cite{DonthBuch2001} of Levy
defects for the main transition ($mt=a$ plus $\alpha $ process) and for the
Fischer modes ($\phi $).\newline
\newline

\begin{figure}[here]
\begin{center}
\includegraphics[width=0.5\linewidth]{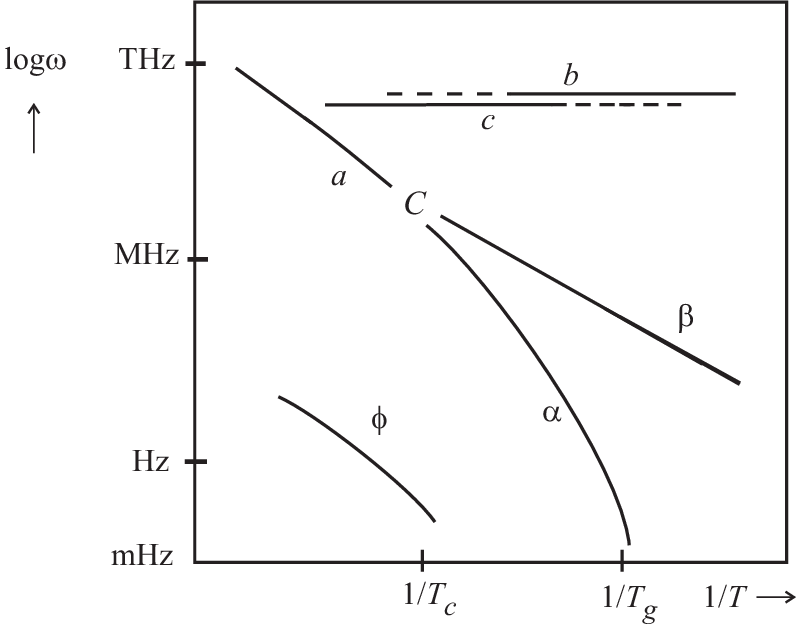}
\caption{Relaxation chart for moderate liquids; $\log \omega $ the
mobility, $T$ the temperature. Dispersion zones: $a$ the
high-temperature process, $\alpha $ the cooperative process,
($mt=a$ plus $\alpha =$ the main transition or main process = The
dynamic glass transition), $\beta $ the
local Johari Goldstein process, $C$ the $a-\alpha $ crossover region, and $%
\phi $ the Fischer modes. [Boson peak $b$ and cage rattling $c$
are not further discussed.] The conventional glass temperature
$T_g$ corresponds approximately to the intersection of the $\alpha
$ process with the experimental 10~millihertz isochron. $T_c$ is
the crossover temperature. More complicated substances such as
polymers or liquid crystals may have additional dispersion zones.
\label{fig3}}
\end{center}
\end{figure}

The concentration of mobility ($\log \omega $) into zones can be explained
by the defects. The width of the imaginary part of the susceptibility
(dynamic compliance $\alpha ^{\prime \prime }(\omega )$) is, because of $%
\alpha ^{\prime \prime }(\omega )\sim \omega x^2(\omega )$, determined by
the embracing Levy distribution for $x^2(\omega )$. The corresponding $%
\alpha ^{\prime \prime }(\omega )$ graphs for $\alpha \leq 1$ show (Ref.~%
\cite{DonthBuch2001}, p. 307 and p. 311) an approximate half width of one
decade in frequency $\omega $, divided by the Levy exponent $\alpha $,

\begin{equation}
\Delta \log \omega =(1.05\pm 0.05)\,/\,\alpha .  \label{Eq.6.1}
\end{equation}
[This equation is valid independently on the spatial dimension of the defect
model. The width of the $\alpha ^{\prime \prime }(\log \omega )$ peak
depends only from the Levy exponent $\alpha $.] Large mobility distances
between the zones in the relaxation chart care for large attempt rates in
partial systems as used for the above inside treatment.\newline
\newline

The non-Arrhenius behavior of the main transition $mt$ is, in principle,
connected with the increase of its cooperativity $N_{mt}$ at low temperature
(Fig.~6, below) via a Levy scaling equation

\begin{equation}
d\log \omega \,/\,dT\sim N_{mt}^{1/\alpha },  \label{Eq.6.2}
\end{equation}
where $N_{mt}$ is the number of molecular units in the $mt$ defect. This
equation is related to the existence of a bulk modulus $m_f$ being the
''reciprocal'' of an invariant free volume $v_f\sim 1/m_f$,

\begin{equation}
v_f=dV^{\prime }\,/\,d\log \omega ,  \label{Eq.6.3}
\end{equation}
invariant against the partition of a larger system into partial systems
(Section~VI.D).\newline
\newline

A Levy treatment of partial systems inside one minimal subsystem (Ref.~\cite%
{DonthBuch2001}, p. 327, see also Section~VI.D, below) is explicitly based
on our main assumption to consider spatially separable components of a Levy
sum. Further development \cite{Donth1996} leads to an extreme smallness of $%
m_f$ for small Levy exponents $\alpha $ at low temperature, $m_f$ ($\alpha
\lesssim 0.4)=o(m_f(\alpha \approx 1))\approx m_f(T=T_c)$. The large
curvature of the $mt$ in the relaxation chart follows from (\ref{Eq.6.2})
with a thermodynamic equation along the $mt$, $dV^{\prime }\sim dT$: $d\log
\omega /dT\sim d\log \omega /dV^{\prime }\sim 1/v_f\sim m_f$ increases more
than Arrhenius, $d\log \omega /dT\sim 1/T^2$. The equation

\begin{equation}
m_f\sim v_f^{-1}\sim N_{mt}^{-1/\alpha }  \label{Eq.6.4}
\end{equation}
is called {\it control equation}. This actual operational freedom $m_f$
becomes narrow inside large units, irrespective of the dictatorial control
by a preponderant component.\newline
\newline

Remark. The extreme smallness of the operational freedom $m_f$ below $\alpha
\lesssim 0.4$ implies an exhaustion of the $mt$ process at low temperature
where, in the equilibrium, not enough free volume can be organized by
increasing cooperativity of the defect periphery. The $mt$ process is then
successfully ''attacked'' from distributed ''accidental'' loss effects (Ref.~%
\cite{DonthBuch2001}, p. 178), and the control by equation (\ref{Eq.6.4}) is
expected to be lost at a surprisingly sharp \cite{Donth1996} exhaustion
transition.\newline
\newline

\subsection{Gedankenexperiment for finding a minimal subsystem}

If external large heat reservoirs and the imagination of walls between
subsystems are given up and if, for dynamics, the subsystems are defined by
statistical independence in a given dispersion zone, then the larger
subsystems are free in partition, and all subsystems are freely fluctuating
in all relevant thermodynamic variables. We obtain freely fluctuating
subsystems.\newline
\newline

The following gedankenexperiment is for the definition of
representativeness. Consider a sufficiently large freely fluctuating
subsystem having a certain spectral density $x^2(\omega )$ (e.g. for
fluctuating (free) volume in the main transition and the following
increments for thermodynamic variables across the $mt$). Divide this
subsystem into two halves. If they also are sufficiently large, then both
have the same non-extensive (e.g. density-like) variables. It is said that
they are {\it representative}. Divide again, till the subsystems become too
small for representativeness. We find finally a minimal representative
freely fluctuating subsystem for the given dispersion zone: shortly a {\it %
minimal subsystem}.\newline
\newline

Representative and minimal means that this subsystem contains one defect and
has a size of order the defect size (Fig.~2d). The defect labels the minimal
subsystem as a thermodynamic unit. The unit has three corresponding parts:
preponderant component, center, and periphery, united by a Levy
distribution. [Since the only true unity of the defect is the one
preponderant component, we may say that this component makes the minimal
subsystem to a thermodynamic unit: to ''the molecule'' of $mt$ fluctuation
of the liquid.]\newline
\newline

Representative subsystems have the same relevant thermodynamic variables as
the macroscopic systems, with all its fluctuations. This means that also the
minimal subsystem has a temperature $T$ that can completely be defined via
Carnot and Kelvin, and has an entropy $S$ that can completely be defined via
Clausius; both $T$ and $S$ fluctuate, $\delta T\neq 0$, $\delta S\neq 0$,
also for the minimal subsystem when it is in the nanometer range. The
systems from the gedankenexperiment plus arbitrary combinations of them are
expected to be candidates for equivalent and robust components in an
appropriate Levy sum for thermodynamics of liquids.\newline
\newline

\subsection{Representativeness Theorem. Pluralism}

We conclude from the above gedankenexperiment that the shape of spectral
density $x^2(\omega )$ in all representative subsystems is the same. We
expect additionally from the equivalence of all partitions into a sum of
representative subsystems that for additive, i.e. extensive variables the
shape of $x^2(\omega )$ is fixed to be a density of a Levy distribution. We
show, so to speak, that the bare existence of minimal subsystems in liquids
implies the defect. As an example, we think firstly on $x^2(\omega )=\Delta
V^{\prime 2}(\omega )$ as a density for free volume ($V^{\prime }$)
fluctuations in the main transition ($mt$).\newline
\newline

{\bf Representativeness theorem}. (Ref.~\cite{DonthBuch2001}, p. 237).
Consider the classical (non-quantum mechanical = symmetrical) spectral
density for stationary fluctuations of free volume, or a corresponding
extensive caloric variable, in the slower part of the $mt$ dispersion zone ($%
\log \omega $ (rad/s) $<11$) of a representative freely fluctuating
subsystem. This spectral density is a Levy distribution density $f(x)dx$
with frequency measure $dx=d\omega $ and a Levy exponent $\alpha \leq 1$.
The exponent $\alpha $ depends on temperature $T$, pressure $p$, substance,
and kind of response.\newline
\newline

{\bf Remark}. The appropriate variables for measurement are
susceptibilities, e.g. dynamic ones as function of frequencies ($\alpha
^{*}(\omega )$ according to Eq.~(\ref{Eq.1.3}); i.e. moduli $m(\omega )$ for
intensive and compliances $j(\omega )$ for extensive thermodynamic
variables). The thermodynamic increments for e.g. the extensive (free)
volume follow from integrals across the dispersion zone (DZ),

\begin{equation}
\Delta V=\int\limits_{\text{DZ}}V^2(\omega )\,d\omega .  \label{Eq.6.4a}
\end{equation}
Defining the volume compliance by $B=-\partial V/\partial p$, [$B$] = m$^3$%
/Pa, we get for the FDT (the ''measuring equation'' \cite{DonthBuch2001}
p.269 ff., \cite{DonthWZ1982})

\begin{equation}
\overline{\Delta V^2}=k_BT\,B,  \label{Eq.6.4b}
\end{equation}
with a correlation function $\Delta V^2(t)$ in the time domain,

\begin{equation}
\Delta V^2(t)=k_BT~(B(t)-B_{\text{equil}}),  \label{Eq.6.4c}
\end{equation}
or with a spectral density $\Delta V^2(\omega )$ in the frequency domain,

\begin{equation}
\Delta V^2(\omega )=k_BT\,B^{\prime \prime }(\omega )\,/\,\pi \omega ,
\label{Eq.6.4d}
\end{equation}
with $B^{*}(\omega )=B^{\prime }(\omega )-iB^{\prime \prime }(\omega )$,
real part minus imaginary part.\newline
\newline

{\bf Proof}. The proof is divided in five steps.\newline
\newline

1. Let us recall some terminological foundations of probability theory.
Consider the random variable ${\bf X}_1=V_1^{\prime }$. The observable, $\nu
_1>0$ or $v_1>0$ for $V_1^{\prime }$, means a coordinate in the sample space
$R^1$, i.e. a variable for a ''conceptual experiment'' of probability. [For
thermodynamics, this experiment is described by the FDT]. The distribution $%
F_1(v_1)$ for the random variable is defined by the probability $P$ to
observe a value $\nu _1<v_1$, $F_1(v_1)=P[\nu _1<v_1]$; the corresponding
density is denoted by $f(v)$, $P[a<v<b]=\int\limits_a^bf_1(v_1)dv_1$.
Analogous definitions are used for the other subsystems, \{$V_2^{\prime
},...,V_i^{\prime },...V_n^{\prime }$\}.\newline
\newline

For the sum of independent free volumes $V_i^{\prime }$ of two subsystems, $%
S_2^{\prime }=V_1^{\prime }+V_2^{\prime }$, we have $F_{S2}(s_2)$ with a
convolution (\ref{Eq.2.3}) for the density,

\begin{equation}
f_{S2}(s_2)=\int f_1(s_2-v)~f_2(v)\,dv=f_1*f_2(s_2).  \label{Eq.6.5}
\end{equation}
The variance and expectation for any free volume exist, e.g.

\begin{equation}
E(s_2)=\int s_2\,f_{s2}(s_2)\,ds_2.  \label{Eq.6.6}
\end{equation}
The distribution for $S_n^{\prime }$ as function of the variable $s_n$
tends, therefore, to a Gauss limit distribution. In general, we have a Gauss
bell curve for $f_V(v)=f(v)$.\newline
\newline

2. Interested in dynamics of liquids, we must change the variables from
volume to frequency, $v\rightarrow v(\omega )$. Consider the representative
inverse function $\omega (v)$ of Fig.~1, e.g. via the inside treatment of
the local breakthrough of mobility, our Levy instability for $\alpha <2$.
Practically, substituting $\omega =\Omega \approx 10^{12}$rad/s by $\omega
=\infty $, we find a continuous function diverging at some finite, maximally
accessible free volume inside the minimal subsystem, $V_{\max }^{\prime
}<\infty $:

\begin{equation}
\omega (v)\rightarrow \infty \text{\quad for\quad }v=v_{\max }.
\label{Eq.6.7}
\end{equation}
The expectation value of frequency may therefore diverge,

\begin{equation}
E(\omega )=E(\omega (v))=\int \omega (v)\,f(v)\,dv\rightarrow \infty ,
\label{Eq.6.8}
\end{equation}
and we expect a new distribution, $f_V(\omega )=$ $f(\omega )$, different
from Gauss.\newline
\newline

One may ask, whether the general shape of the new, dynamic density $f(\omega
)$ depends on details of the $\omega (v)$ function. The answer is no, since
the convolution property (\ref{Eq.6.5}) for many components $v_i$ pushes the
distribution for their sum into the domain of attraction of a limit
distribution. Then from a divergence (\ref{Eq.6.8}) we expect a Levy
distribution with Levy exponent $\alpha \leq 1$, whatever the details of the
continuous $\omega (v)$ function beyond the property (\ref{Eq.6.8}) are. The
details of $\omega (v)$ can, however, influence the particular value of $%
\alpha $. It is interesting that changing the measure may imply a change of
the exponent $\alpha $ (cf. the Corollary).\newline
\newline

3. What is the appropriate (physical) density function $f(\omega )$ for the
stationary fluctuation of an extensive variable? This is the spectral
density, $f(\omega )=$ $x^2(\omega )$ (\ref{Eq.1.2}), because the
corresponding variable measures are equivalent as indicated by the integral
Eq.~(\ref{Eq.6.4a}),

\begin{equation}
\int x^2(\omega )\,d\omega =\int f(v)\,dv=\Delta V_{mt}^{\prime },
\label{Eq.6.9}
\end{equation}
where $\Delta V_{mt}^{\prime }$ is the (free) volume step increment for
crossing the main transition ($mt$). The ''additivity'' of $x^2(\omega )$ is
also ensured by its equivalence to the compliance $\partial V/\partial p$ or
$\partial V/\partial T$ via the FDT.\newline
\newline

4. From (\ref{Eq.6.5}) for free volume we get, after the substitution $%
f(v)\rightarrow f(\omega )=x^2(\omega )$ for each component of the Levy sum,

\begin{equation}
X_s^2(\omega )=(X_1^2*X_2^2)(\omega ).  \label{Eq.6.10}
\end{equation}
Hence, for general $n$, we have for any sum of minimal subsystems

\begin{equation}
X_{sn}^2(\omega )=X_1^2*...*X_n^2(\omega ).  \label{Eq.6.11}
\end{equation}
Physically, the general frequency variable $\omega $ follows from the $%
\omega $-identity of the FDT, measuring a general frequency for any set of
representative subsystems and the same frequency $\omega $ for any activity,

\begin{equation}
\{\omega _i\}\rightarrow \omega .  \label{Eq.6.12}
\end{equation}
Since (\ref{Eq.6.11}) is a sufficient condition for a Levy distribution, we
get such a distribution for $x^2(\omega )$ with measure $d\omega $.\newline
\newline

In the Nyquist model~\cite{DonthBuch2001,Nyquist1928} for the FDT (mentioned
in Section III.A), the $\omega $-identity is related to the
indistinguishability of the $\hbar \omega $ quantums in the transmission
lines between ''sample'' and ''apparatus'': different activities correspond
to different selections of quantums from these lines.\newline
\newline

5. A thermodynamic argument for exponent $\alpha <1$ follows from the
periphery of the defect. The Levy distribution connects the center of the
defect (fractality) with the periphery (tail of stretched exponential,
Fig.~2c). The long-time retardation is in the periphery. The simplest case
for retardation is a ''thermodynamic'' or ''quasi-stationary'' one (in the
terminology of Landau and Lifshitz, Ref.~\cite{LLBuch1980}, \S 118)

\begin{equation}
dx/dt\sim -x\quad ,\quad x\rightarrow 0\quad \text{for\quad }t\rightarrow
\infty .  \label{Eq.6.13}
\end{equation}
This gives a Debye (exponential) decay corresponding to a Cauchy
distribution (\ref{Eq.2.4}) ($\alpha =1$) in the frequency domain.\newline
\newline

The cooperativity of the periphery induces a spectrum of such thermodynamic
decays. This means stretching the retardation. For Levy distribution we get
necessarily the Kohlrausch function (\ref{Eq.1.1}), i.e. a Levy exponent $%
\alpha <1$, holding for the total defect because of the correspondence
between center and periphery. This ends the Proof, if the question of
variables wherefrom the exponent $\alpha $ may depend (and physically
depends) is delegated to the

{\bf Corollary}. The only free parameter in the correlation function for
fluctuations of extensive variables in the main transition of moderate
classical liquids is the Levy exponent $\alpha \leq 1$. The exponent $\alpha
$ is representative, but remains non-fixed for minimal subsystems in the
one-nanometer range. It is therefore expected to be influenced by the
structure and the responsibility (Fig.~2c) of the molecules and of the
molecular structure of the Levy defect. This includes the dependence on the
thermodynamic state (temperature $T$, pressure $p$, composition $x$,...). We
have got

\begin{equation}
\alpha =\alpha (\text{activity, substance, }T,p,x,...).  \label{Eq.6.14}
\end{equation}
Activity means the different kind of response for extensive variables, e.g.
volume, entropy, dielectric polarizability, incoherent or coherent dynamic
neutron scattering, and so on.\newline
\newline

Six Remarks. (1) The Representativeness Theorem shows that in moderate
liquids the outside approach to the Levy distribution via representative
subsystems is consistent with the inside approach via partial systems.%
\newline
\newline

(2) The Gauss distribution for free volume from Eq.~(\ref{Eq.6.2}) excludes
the picture, that there is a partial system with a preponderant contribution
of the free volume itself. But this Gauss distribution does not exclude a
spatial concentration of such partial systems that do have more free volume
than others. The general space/time scaling (Fig.~2) can generate a defect
with more free volume from these partial systems in the center.\newline
\newline

(3) Does the correspondence between center and periphery mean that there is
no diffusion step through the cage door in a Levy liquid without a
preponderant component?\newline
\newline

(4) Long range correlation as from lattices in solid state physics and long
range quantum correlations in liquids may prevent a reasonable partition
into smaller and smaller independent subsystems, so that representativeness
does not longer remain a reasonable concept: No spatial Levy defects there.%
\newline
\newline

(5) In other words, the Corollary expresses some kind of {\it pluralism} of
different activities. Although the different activities are collected by the
molecules, in a way, in the same dispersion zone, and although always the
same molecules of the minimal subsystem are involved, the activities are
different in shape and location across the zone. This is related not only to
the value of the Levy exponent $\alpha $, but also to modifications from
molecular particularities in the center (e.g. proximity of NMR signals to
mechanics) or in the periphery (e.g. nonlocal properties of shear response).
Each activity for itself is related to Levy, but each has different Levy
exponents and possible modifications, even in the same ($p,T$) state. The
shape of the limit distribution has different ''faces''. Although frequently
confirmed by experiments, such a pluralism of a theoretical approach seems
difficult to be accepted by the glass transition community.\newline
\newline

(6) A drastic example for the pluralism is the above variable-change from
free volume $v^{\prime }$ to frequency, $\omega (v^{\prime })$. For $%
f(v)=f(v^{\prime })$ we get a Gauss distribution ($\alpha =2$), and for the
spectral density $f(\omega )=x^2(\omega )$ we get a Levy distribution with $%
\alpha <1$. Such findings may have consequences on the statistical physics
in liquids (Section~VII, von Laue approach). The problem is hidden, because
Gauss behaviour ($\alpha =2$) may be expected for all thermodynamic
potentials that may be found in the nominator of Gibbs exponents, from
similar arguments as for free volume.\newline
\newline

\subsection{Derivation of invariant free volume and control equation}

The Levy treatment of Section VI.A is based on an inside treatment of
cooperativity, inside one minimal subsystem i.e. inside one defect
(Section~VI.B). Consider independent partial systems with a fixed average
volume. The size of the subsystem is then proportional to the number $m$ of
such partial systems, $N_{mt}\sim m$. Consider the frequency $\omega _i$, $%
i=1,...,m$, as a transition probability. Statistical independence is then
expressed by a common product,

\begin{equation}
\omega _g=\text{const}\times \omega _1\times \omega _2\times ...\times
\omega _m.  \label{Eq.6.3a}
\end{equation}
From (\ref{Eq.3.1}) and Fig.~1, we have

\begin{equation}
\omega _i=\omega _i(V_i^{\prime })\text{ , same }i~;  \label{Eq.6.3b}
\end{equation}
from additivity of free volume $V^{\prime }$ we have

\begin{equation}
V^{\prime }(\text{subsystem) }=V_1^{\prime }+V_2^{\prime }+...+V_m^{\prime }.
\label{Eq.6.3c}
\end{equation}
These three equations define a functional equation yielding the general
solution

\begin{equation}
\omega (\text{subsystem) }=\omega _M\exp (\frac{V_1^{\prime }+V_2^{\prime
}+...+V_m^{\prime }}{v_f})  \label{Eq.6.3d}
\end{equation}
with $\omega _M$ a prefactor depending of the partition and $v_f$ a scaling
parameter of the minimal subsystem not depending on the partition. This $v_f$
is called the invariant free volume. The cooperativity is now described by
exchange of free volume between the independent partial systems balanced by
the relatively small fractional volume fluctuation $\delta V/V$ of the total
minimal subsystem (''minimal coupling'').\newline
\newline

Intuitively, the decrease of operational freedom $m_f$ (Section~VI.A) is
connected with smaller free volume. Comparing directly the scaled Levy sum $%
\sum {\bf X}_i/n^{1/\alpha }$ (Eq.~(\ref{Eq.2.1a}), cf. also Theorem 1,
above) with the exponent in Eq.~(\ref{Eq.6.3d}), we would get $v_f\sim
N_{mt}^{1/\alpha }$, i.e. $v_f$ would increase with the cooperativity. This
means that $v_f$ cannot be directly proportional to this freedom. To analyse
the situation (free volumes $V_i^{\prime }$ also decrease with lower
temperatures) we are led to susceptibilities near equilibrium, being the
direct measures from the FDT (Section~VI.C). We have then dynamic
compliances $j(\omega )$ that become larger for softer materials (high
temperature, small cooperativity $N$), and moduli $m(\omega )$ that become
larger for harder materials (low temperature, large cooperativity). That is,
the operational freedom, decreasing for low temperatures, increases with the
modulus. Using the rheological equation $j(\omega )\cdot m(\omega )=1$, and
putting $j\sim v_f\sim N_{mt}^{1/\alpha }$, we get for the operational
freedom $m_f\sim m(\omega )\sim 1/j\sim v_f^{-1}$, i.e. really Eq.~(\ref%
{Eq.6.4}). [In the book \cite{DonthBuch2001} and Ref \cite{Donth1996} this
relation was used without any derivation.] Since the general Levy scaling by
$n^{1/\alpha }$ is for Gauss consistent with the fluctuation variance (\cite%
{FellerBuch}, p.172), the scaled Levy sum seems to be the only possibility
to get a general damping of the operational freedom with increasing
cooperativity $N$. Reducing with Gauss ($\alpha =2$, $m_f\sim N^{-1/2}$), we
get $m_r=m_f/m_f($Gauss$)\sim N_{mt}^{(\alpha -2)/2\alpha }$, e.g. $m_r\sim
N_{mt}^{-3/2}$ for $\alpha =1/2$, i.e. a considerably stronger restriction
of freedom than from Gauss.\newline
\newline

\subsection{Johari Goldstein $\protect\beta $ process}

Consider falling temperatures. The $\beta $ process ''emerges'' in the $a$-$%
\alpha $ process crossover region C of the relaxation chart Fig.~3.
According to the idealized version of the mode coupling theory MCT \cite%
{Goetze1992}, the cage door of next molecular neighbors (for the $a$
process), necessary for the diffusion in liquids, is closed at the critical
temperature $T_c$ of this MCT. In our Levy defect approach, $T_c$ is
substituted with the crossover temperature: Because of the generality of
defects, esp. some continuity of molecular motion in the cooperative
periphery (see Fig.~6, below), Johari's islands of mobility (Section III
B,C) survive the crossover with a modification \cite{Stickel95+96}. The
increasing cooperativity of the $\alpha $ process, located in the periphery
of the defect, leaves the cage door open for diffusion. The position of the
crossover region in the relaxation chart is then determined by crossing of
the $a$ process with a virtual molecular-mechanical process for $\beta $,
finding free room enough in the center of the defect. From this point of
view, the parameter change for the $mt$ at the crossover \cite{Stickel95+96}
means a change from a defect without ($T>T_c$) to a defect with the inside
mechanical process ($T<T_c$). [The parameters of the $a$ process and the
mechanical core of the $\beta $ process must suffice some relation to get a
crossing. There are glass formers where no crossing is observed, i.e. with
''no $\beta $ process'' \cite{Kudlik1999}].\newline
\newline

A $\beta $ process is approximately of Arrhenius type and is accompanied by
a ''continuous'' background from the periphery. The total formal variance of
the $\beta $ process can be affected (i.e. broadened) by a distribution of
molecular activation-barrier properties. A concrete $\beta $ activity can be
rather small, if the mechanical process does not generate the corresponding
fluctuations.\newline
\newline

The main issue is the relation of the $\beta $ process to the $\alpha $
process below the crossover region. Wherefrom does the free volume
concentration necessary for the above activation process come? Although the
free volume of the bulk becomes extremely small at low temperatures, so that
the $\alpha $ process needs more and more cooperativity for perpetuating the
diffusion steps through the cage door? A short answer is the phase
continuity of the liquid state so that cooperativity must increase to ensure
the fluidity of the liquid, even at the cost of long times (large viscosity)
at low temperature, cf. Eq.~(\ref{Eq.6.2}). The longer answer from our
approach is that free volume enough can be collected in the center of Levy
defects of the cooperative $\alpha $ process (with the preponderant
component for dynamics in the midst). The correspondence between center and
periphery then guarantees that increasing cooperativity of the periphery for
falling temperatures (Fig.~6, below) ensures the continuous existence of the
mechanical process by the preponderant component for $\alpha <1$, also when
the total free volume becomes shorter. [Let us repeat: In the idealizing MCT
closing of the cage door is possible; in our approach the periphery cannot
end its action to keep the cage door open.] The alternative \cite%
{NgaiPaluch2004}, that the $\beta $ process is a precursor or local step of
the $\alpha $ process which can occur anywhere \cite{NgaiPC2005}, cannot
answer the above wherefrom question. In our defect picture, the anywhere
would only be right in the sense that the Levy defect can occur anywhere in
the pattern, but anywhere outside the defect center would not be probable.%
\newline
\newline

\subsection{Fischer speckles}

One of the greatest surprise in glass transition research was the discovery,
that the Debye Bueche \cite{Debye1949} inhomogeneities in frozen glasses
correspond to a slow dispersion zone $\phi $ in the relaxation chart Fig.~3.
These Fischer modes have e.g. times about 10$^7$ times longer and lengths
about 50 times larger than those of the main transition process ($mt$) \cite%
{Fischer1993}. It follows from the Representativeness Theorem that there
must be also defects in the fluctuating pattern for the $\phi $ process
zone. These defects are to be identified with the experimental \cite%
{Patkowski2001+Bakai2004} Fischer speckles.\newline
\newline

The volumes of partial systems for the inside approach to the Glarum defects
of the main transition are smaller than the volumes of the $mt$ minimal
subsystems, whereas for the Fischer speckles the partition into just such
subsystems (as ''partial systems'' for $\phi $) is used (Fig.~4).\newline
\newline

\begin{figure}[here]
\begin{center}
\includegraphics[width=0.5\linewidth]{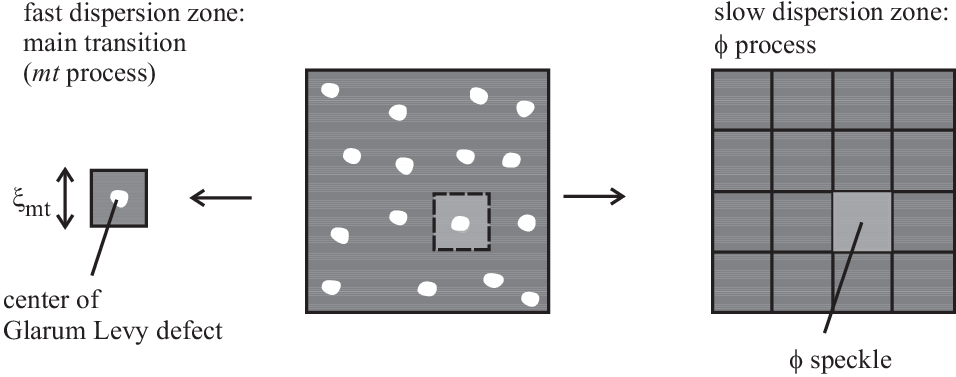}
\caption{Fischer speckles ($\phi $ speckles). Two Levy situations
in moderate liquids. Left for the main transition $mt$, right for
Fischer modes $\phi $. The small squares are the minimal
subsystems for the main transition. \label{fig4}}
\end{center}
\end{figure}

The Fischer modes $\phi $ are a direct consequence of the main transition $%
mt $. No structural thermodynamic assistance (''cluster'') is necessary for
the construction of the length and time ratios between $\phi $ and $mt$
processes (Fig.~5). These ratios can be estimated from alone the two-defect
pattern of Fig.~4. We have to look for a physical $mt$ process in the liquid
that eats its way through large space $\lambda $ and time $t$ ranges:
diffusion process. How probes the fast $mt$ process the slow $\phi $ process
speckles? What about the reliability of the short-range $mt$ process for the
long-range $\phi $ process?\newline
\newline

\begin{figure}[here]
\begin{center}
\includegraphics[width=0.5\linewidth]{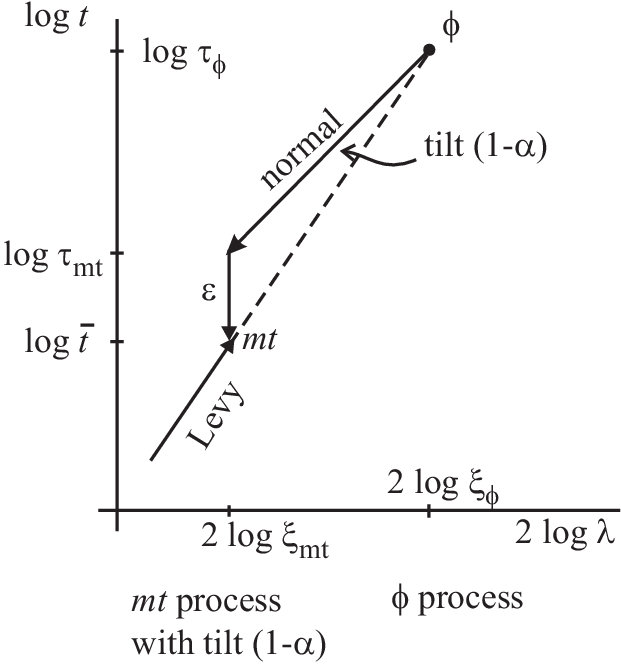}
\caption{Tilt construction for Fischer $\phi $ process length
($\xi _\phi $) and time ($\tau _\phi $) in a diffusion plot, $\log
t$ vs $\log \lambda ^2$, using the relevant tilt ($1-\alpha $) of
the Glarum Levy defects of the main transition ($mt$).
\label{fig5}}
\end{center}
\end{figure}

The $mt$ process defects are rather dense, their average distance is of
order 1~nm (see Fig.~6, below). The molecular diffusion is therefore step by
step, from one Glarum Levy defect to the next one along the ''diffusion
line''. We expect locally, at any step, a modification of the ''normal
diffusion'' by the Levy fractality at the cage doors: ''Levy diffusion''.
This defines the local $\log t/\log \lambda $ slope, and an ''analytical
continuation'', step by step along the line gives the required reliability
of the defect behavior over the full distance between the speckles. The
Representativeness Theorem includes also the large subsystems between the
Fischer speckles. $-$ Conversely, the fast $mt$ process should also be
''probed'' by components of the slow $\phi $ process. We have got two
aspects:\newline
\newline

Seen from the fast $mt$ process. For instance, dynamic neutron scattering %
\cite{Arbe1998,Arbe2003} shows that the $a$ process can be described by a
sublinear diffusion whose exponent may be related to the Kohlrausch (Levy)
exponent $\alpha $ of a certain $a$ process relaxation. In the frame of
pluralism (Section~VI.C), this exponent may be different from a dielectric
exponent (cf. Eq.~(\ref{Eq.6.14})) \cite{Tyagi2005}. The corresponding Levy
diffusion is described by a slope from

\begin{equation}
\omega \rightarrow \omega ^\alpha ~,~t\sim \lambda ^{2/\alpha }.
\label{Eq.6.15}
\end{equation}

Seen from the slow $\phi $ process. Consider the flaring up (1) and fading
away (0) of the $\phi $ speckles at different space and time ''points'' as
some kind of (1,0) spin diffusion from virtual, spatially fixed spin
carriers. Since different minimal subsystems for the $mt$ process are
independent, this spin diffusion is normal:

\begin{equation}
t\sim \lambda ^2.  \label{Eq.6.16}
\end{equation}

We estimate the required $\phi /mt$ ratios from the triangle in Fig.~5. We
find the nearest neighbors of $\phi $ process defects (speckles) at the
upper $\phi $ corner of the triangle, since the Levy diffusion, to be
effective, must be faster than the normal diffusion. The basis $\varepsilon $
is the log-time difference $\varepsilon $ between normal and Levy diffusion
at the $mt$ process length scale, $\lambda \approx \xi _{mt}$. The Levy
diffusion is at some time $\bar t$ in the short time tail (fractality) of
the $mt$ process spectral density, i.e. in the molecular cage door being the
center of the Glarum Levy defect. The normal diffusion is near the average
time of the $mt$ process, $t_{av}$:

\begin{equation}
\varepsilon =\log (t_{av}/\bar t)_{mt}.  \label{Eq.6.17}
\end{equation}

From Fig.~5 we see that the time and the length ratios ($\phi /mt$) are
mainly determined by the tilt ($1-\alpha $). Defining $\tau _{mt}=t_{av}$,
we get the slope difference as ($1-\alpha )/\alpha $ and

\begin{equation}
\log \frac{\xi _\phi }{\xi _{mt}}=\frac \alpha {2(1-\alpha )}\varepsilon
~,~\log \frac{\tau _\phi }{\tau _{mt}}=\frac \alpha {1-\alpha }\varepsilon .
\label{Eq.6.18}
\end{equation}
For small tilts, $1-\alpha \ll 1$, we obtain the slope difference directly
as the tilt ($1-\alpha $), and $\log (\xi _\phi /\xi _{mt})\approx
\varepsilon /2(1-\alpha )$ as well as $\log (\tau _\phi /\tau _{mt})\approx
\varepsilon /(1-\alpha )$. Large values for the $\phi /mt$ ratios can easily
be obtained, also in the experimental range for moderate molecular liquids
(Ref.~\cite{DonthBuch2001}, p. 349). A minimal subsystem for the $\phi $
process as claimed by the Fischer speckles contains a large number of Glarum
Levy defects.\newline
\newline

\section{von Laue Thermodynamics for liquids?}

\subsection{Characteristic length of the main transition}

The characteristic length is a concept for an estimation of the defect size
from thermodynamic bulk measurements (Fig.~2(d), Sections III.D and VI.C).
This length is decisive for an Experimentum Crucis of the defect pattern.%
\newline
\newline

The usually applied molecular Gibbs statistics rests on a virtual transfer
of the temperature $T$ from a macroscopic heat reservoir, transferred
without temperature fluctuation to the subsystems, $\delta T\equiv 0$. This
transfer does not seem reasonable, if one tries to eliminate external
experiments (the conscious observer of the Copenhagen program) from
thermodynamics or quantum theory of independent subsystems.\newline
\newline

Statistical independence of Landau subsystems is rested on negligible
potential energy between the foreign molecules according to the Boltzmann
factor $\exp (-E_{\text{pot}}(p,q)/k_BT)$ in the Gibbs probability. This
does not seem appropriate for dynamics that is based on large attempt rates
allowing much smaller independent subsystems.\newline
\newline

It seems more reliable to substantiate a thermodynamics of liquids by time
or mobility integration of dynamic compliances and dynamic moduli from the
fluctuation dissipation theorem (FDT) that can be derivated from quantum
mechanics without external experiments (Ref.~\cite{DonthBuch2001}, p. 227ff, %
\cite{Nyquist1928,DonthWZ1982}, cf. the Remark in Sec. VI.C). This leads,
for ($T,S;~p,V$) as relevant thermodynamic variables, to the von Laue
thermodynamics including temperature fluctuation $\delta T$ also for the
above minimal subsystems in the one-nanometer range (Refs.~\cite%
{vonLaue1917,LLBuch1980} \S 112, and \cite{Kondepudi1998} Chap. 14).\newline
\newline

A thermodynamics based on the dynamic compliances ($\alpha ^{\prime \prime
}(\omega )\sim \omega x^2(\omega ))-$ without any restriction of fluctuation
of thermodynamic variables, similar to the von Laue thermodynamics $-$ may
possibly be associated with a statistical basis different from the more
''static'' Gibbs distribution. In particular, the separation into mechanics (%
$E(p,q)$) and thermodynamics ($k_BT$) of the Gibbs distribution may be lost.
Possibly, there exists a molecular, microscopic von Laue distribution with
different consequences near the defects, related to the Levy distribution as
representative of minimal subsystems and reflecting the pluralisms
(different $\alpha $ for different activities) more explicitly.\newline
\newline

From the von Laue fluctuation formulas we get directly \cite{DonthJNCS1982}a
volume that can be interpreted as volume of the minimal subsystem when
applied to the main transition $mt$. We start from one defect as the
smallest representative thermodynamic unit (Section VI.C). In the general
von Laue fluctuation formula for temperature, $\overline{\Delta T^2}%
=k_BT^2/C_V$, the heat capacity $C_V$ is an extensive variable [J/K]. Having
$N_{mt}$ molecular units in our minimal subsystem, we get

\begin{equation}
N_{mt}\text{ (von Laue) }=\frac{RT^2\Delta \,(1/c_V)_{mt}}{M_0\,(\delta
T)_{mt}^2},  \label{Eq.7.1}
\end{equation}
where now the small $c_V$ letter is the specific heat capacity [J/kg$\cdot $%
K]: $\Delta (1/c_V)_{mt}$ is the step of $1/c_V$ across the $mt$, $R$ is the
molar gas constant, and $M_0$ is the molecular mass of what is considered as
molecular unit. The characteristic length is obtained from the average
volume needed by the $N_{mt}$ molecular units; $N_{mt}$ was called ''the
cooperativity'' for the defect of the $mt$.\newline
\newline

We discuss now the extensivity (being the expression of the Levy sum). Put $%
\nu $ the number of minimal subsystems in the sample; i.e. Eq.(\ref{Eq.7.1})
is for $\nu =1$. To get the extensivity of $C_V$, we multiply the heat
capacity of one minimal subsystem with $\nu $ and get then for the
temperature fluctuation of the sample $\overline{\Delta T^2}\sim 1/\nu $, $%
\overline{\Delta T}\sim 1/\sqrt{\nu }$, as expected. But how to get ($\delta
T)_{mt}^2$, the temperature fluctuation of the minimal subsystem, from a
bulk measurement? The Representativeness Theorem makes the relevant
responsibility ($\alpha ^{\prime \prime }(\omega )\sim \omega \Delta
x^2(\omega )$) indepent from the system size, i.e. the width $\Delta \log
\omega $ of the $mt$ dispersion zone (DZ) of Fig.~3 does not become smaller
for large subsystems or samples. We get, across the DZ, a temperature
difference

\begin{equation}
\delta T\approx (dT/d\log \omega )_{\text{along DT}}\cdot \delta \log \omega
.  \label{Eq.7.5}
\end{equation}
To interprete this $\delta T$ as $\delta T_{mt}$, we must assume that for
partial systems of the minimal subsystem, the distributions for temperature
and relevant mobility fluctuations are equivalent. This is a fluctuative
interpretation of Boltzmann's macroscopic temperature-time equivalence used
by rheologists for a long time. [This equivalence is an approximate symmetry
of susceptibility $\alpha =\alpha (\log \omega ,T)$ curves across the $mt$
with respect to a $\log \omega \leftrightarrow T$ exchange, called also
rheological simplicity. Interpreting in this context $\log \omega $ as a
thermodynamic variable equivalent to a temperature $T$, this symmetry can be
thought to be inversely constructed from fluctuative increments for both, $%
\delta \log \omega $ and $\delta T$. This confirms our above assumption, in
particular for relevant susceptibilities $\alpha $, e.g. for temperature
modulus or for entropy compliance $\sim $ dynamic heat capacity, $C(\omega
,T)$. In the FDT as thermodynamic equation for an experiment, this
equivalence is mediated by the Planck radiation formula in Nyquist's
transmission lines, cf. Fig.~3.6b of Ref.~\cite{DonthBuch2001}; a detailed
preliminary discussion of this issue is also in \cite{DonthBuch2001},
pp.263-268. A critique of this argumentation is in \cite{SchroeterJNCS2006}.]%
\newline
\newline

Eq.~(\ref{Eq.7.1}) is from the fluctuation of an intensive variable, the
corresponding susceptibility is a modulus. From a Gibbs fluctuation of
energy, $\left\langle \Delta E^2\right\rangle _V=k_BT^2C_V$ (with the
restriction $\delta T\equiv 0$), we get from an analogous treatment \cite%
{DonthZPC1977,Sillescu1999}

\begin{equation}
N_{mt}\text{ (Gibbs) }=\frac{RT^2}{M_0\,(\Delta c_V)_{mt}\,\delta T_{g,mt}^2}%
.  \label{Eq.7.2}
\end{equation}
This is from the fluctuation of an extensive variable, the susceptibility is
a compliance; $\delta T_{g,mt}$ is the transformation interval after
correction for partial freezing around the thermal glass temperature $T_g$.
The ratio is unexpectedly large, of order

\begin{equation}
\frac{N_{mt}\text{ (Gibbs)}}{N_{mt}\text{ (Laue)}}\approx
{c \overwithdelims() \Delta c}
^2,  \label{Eq.7.3}
\end{equation}
where $\Delta c$ is the step height across the $mt$ of the heat capacity
used for the estimation \cite{Hempel2000}, and $\delta T^2\approx \delta
T_{g,mt}^2$ which assumption is not a point of issue. For e.g. $\Delta
c/c=10 $ per cent we get the ratio of 100 (!)\newline
\newline

\begin{figure}[here]
\begin{center}
\includegraphics[width=0.8\linewidth]{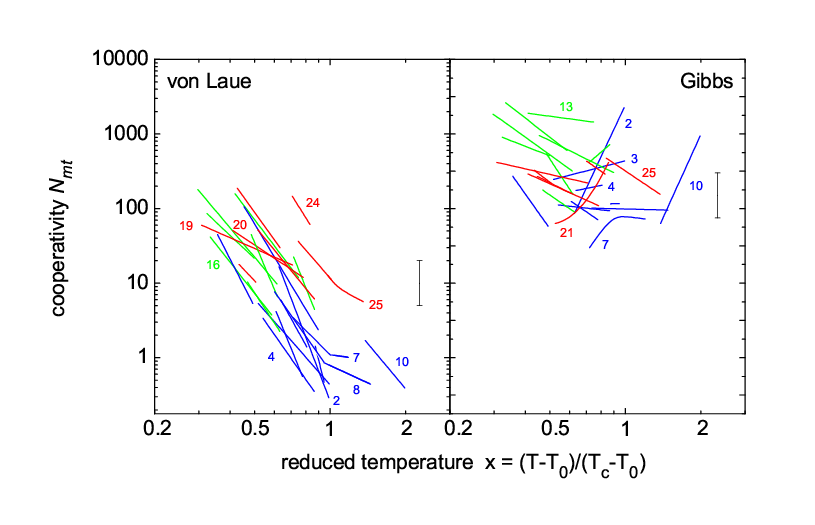}
\caption{Comparison of von Laue and Gibbs cooperativities (Eqs.~(\ref%
{Eq.7.1}) and (\ref{Eq.7.2})) of the main transition $mt$ as a
function of reduced temperature $x$ ($T_c$ the crossover
temperature, $T_0$ the Vogel temperature from a VFT (Vogel Fulcher
Tammann) extrapolation to $\log \omega \rightarrow -\infty $;
$x=1$ means $T=T_c$). The full experimental data are
in Ref.~\cite{HuthDonthSchick}, see also Refs. \cite%
{HuthThesis,DonthEPJE2003} and the references cited therein. The
colors are: blue for poly (n alkyl) methacrylates, green for other
polymers, and red for small-molecule substances. For polymers, the
particles are the monomeric (repeat) units. The numbers are: 2, 3,
7, 8, 10 for poly (n alkyl) methacrylates: 2 - n propyl, 3 - n
butyl, 7 - n pentyl, 8 - n hexyl, 10 - n decyl; 4 - random
copolymer poly (n butyl methacrylate-{\it stat}-styrene) with 2
percent styrene; 13 - polystyrene, 16 polycarbonate, 19 diglycid
ether of bisphenol, 20-poly [(phenyl glycidil ether) - {\it co
}-{\it \ } formaldehyde]; 21 - benzoin isobutyl ether, 24 ortho
cresyl glycidyl ether, and 25 - a fulven derivative (TP-CPBO). The
bars mean an error estimation of the single absolute values.
\label{fig6}}
\end{center}
\end{figure}

The two experimental cooperativities as obtained from dynamic calorimetry
between mHz...kHz frequencies (3$\omega $ method) are compared in Fig.~6 %
\cite{HuthDonthSchick,HuthThesis,DonthEPJE2003}. We see three things.\newline
\newline

(i). The Gibbs cooperativities are much larger than the von Laue
cooperativities. The related Gibbs lengths correspond about to the diameter
of the cage from next molecular neighbors plus parts of the second shell, as
expected for Landau subsystems from the energetically motivated independence
due to the Boltzmann factor.\newline
\newline

(ii). There is no indication for any singularity (cusp or so) at the
crossover temperature \cite{Beiner1998,Kahle1999} (for the substances where $%
T_c$ is included) that could be expected from a critical temperature from
closing the cage of the pristine = idealizing model for mode coupling theory
MCT \cite{Goetze1992,Goetze1984}.\newline
\newline

(iii). The von Laue cooperativities from calorimetry increase continuously
with falling temperature. The increase corresponds to a continuous decrease
of bulk free volume. Some of the cooperativities are, for high temperature
above the crossover, $T\gtrsim T_c$, smaller than one molecular cage from
the next neighbors and suit, therefore, to the extraordinary entropy
fluctuation accompanying the diffusion step through the cage door induced by
the preponderant component in the midst of the defect center. Sometimes, the
cooperativities are of order $N_{mt}\approx 1$: one molecular diameter near $%
T_c$. Some bend in the crossover region is also observed. For low
temperature, the increasing cooperativities seem to indicate the increase of
the defect periphery to collect enough free volume for the center to
maintain the cage door open also for short free volume, as required by the
Representativeness Theorem.\newline
\newline

The experimental characteristic lengths from the von Laue approach are,
therefore, considered as size of the Glarum Levy defects.\newline
\newline

\subsection{Experimentum Crucis}

Is there an Experimentum Crucis that can decide between using von Laue or
Gibbs thermodynamics and, if von Laue is the winner, would indirectly prove
the Levy defects in the main-transition ($mt$) dispersion zone of the
dynamic glass transition (Fig.~3; see also Ref.~\cite{DonthBuch2001} p.187, %
\cite{DonthEPJE2003})? Since the local density effects expected from the
cage doors for diffusion are small due to the steep repulsion potentials of
the particles, it seems at present difficult to detect the defects directly
in the structure factors used e.g. for the evaluation of dynamic neutron or
photon scattering. Additionally, the inter/intra molecular potentials used
for adjustments are also uncertain.\newline
\newline

Therefore, an indirect experimental proof via characteristic lengths is
suggested. Dynamic scattering of relevant contributions from $mt$ process
allows to construct a raster \cite{KahlePC2002} of wave vectors $Q$,
mobilities $\log \omega _0$, and isotherms $T=$~const. Transferring the
maximum frequency $\omega _0^{\text{DC}}$ for the heat capacity peak of $%
C_p^{\prime \prime }(\omega ,T)$ at different temperatures from dynamic
calorimetry (DC) into the $\log \omega _0$-$T$ lines of this raster, the
resulting length $\xi =2\pi /Q$ from scattering can be compared with the
characteristic lengths from the calorimetric cooperativities, Eqs.~(\ref%
{Eq.7.2}) or (\ref{Eq.7.1}). Since their ratio: Gibbs/von Laue is large, in
particular for high temperatures (Fig.~6), even a rough comparison can give
a decision. If we obtain von Laue lengths, then this result is considered as
a certain experimental proof of the defects (see below).\newline
\newline

What about the physical relevance of calorimetry for neutron scattering in
the frame of pluralism (different exponents $\alpha $)? Roughly, by
thermodynamics, scattering corresponds to some density fluctuation, i.e. to
a volume, and calorimetry to an entropy fluctuation. The von Laue
fluctuations are related by thermal expansion: $\overline{\Delta S\Delta V}%
=k_BT(\partial V/\partial T)_P$. This defines a certain relationship between
them. The general basis for comparison, however, is the $\omega $-identity
of the FDT (\cite{DonthBuch2001}, p.279). Take thus dynamic heat capacity
and the diffusion part of dynamic scattering as the two responses that are
to be compared. Then the comparison of maximal frequency $\omega _0$ of the
two compliances Eq.~(\ref{Eq.1.3}) seems reasonable, since the frequencies
of different activities (even for different Levy exponents $\alpha =\alpha _{%
\text{KWW}}$) are identified by the $\omega $-identy. The Kohlrausch
function from the relevant intermediate scattering function, $S^{\text{rel}%
}(Q,t)\sim \exp \{-[t/\tau _{\text{KWW}}(Q)]^\alpha \}$ must be transferred
to a dynamic compliance, $\alpha ^{\prime \prime }(\omega )$. The maximum
frequency $\omega _0$ can then be calculated from Ref.~\cite{DonthBuch2001},
p.306:

\begin{equation}
\ln \,(\omega _0\tau _{\text{KWW}})=0.60607\,(\alpha -1),  \label{Eq.7.4}
\end{equation}
where $\ln $ means the natural logarithm, $\omega _0$ is the angular
frequency in rad/s, with $\omega =2\pi f$, $f$ the frequency in hertz. We
get $\omega _0(Q)$. The reduction to frequencies avoids further comparison
of times from different activities.\newline
\newline

Consider a fictitious example following partly Ref.~\cite{Tyagi2005}
(Fig.~7). From dynamic neutron scattering (DNS) one can separate a relevant
intermediate scattering function $S^{\text{rel}}(Q,t)$ that is informative
about diffusive (Eq.~\ref{Eq.6.15}) and related relaxation processes. This
is one of our response. Calculation of $\omega _0^{\text{DNS}}(Q,T)$ from a
Kohlrausch fit with $\tau _{\text{KWW}}$ as a function of wave vector $Q$
allows to construct a raster of isotherms in an $\log \omega _0^{\text{DNS}}$
vs $\log Q$ plot. This raster allows then to locate the maximum ($\log
\omega _0,T$) pairs from dynamic calorimetry (as the other response) in the
diagram Fig.~7 ($\triangle $) (by using $\omega _0^{\text{DC}}=\omega _0^{%
\text{DNS}}$ from the $\omega $-identity of the FDT, possibly after some
extrapolation of the DNS raster to lower temperatures). Then the abscissa of
the $\triangle $ points gives the $Q$ vector that is to be associated with
dynamic calorimetry, from which the length $\xi =2\pi /Q$ can be compared
with the characteristic lengths from (\ref{Eq.7.1}) and (\ref{Eq.7.2}), ($%
\xi _{mt}^{\text{DC}})^3=N_{mt}\times $ average volume occupied by one
particle.\newline
\newline

\begin{figure}[here]
\begin{center}
\includegraphics[width=0.5\linewidth]{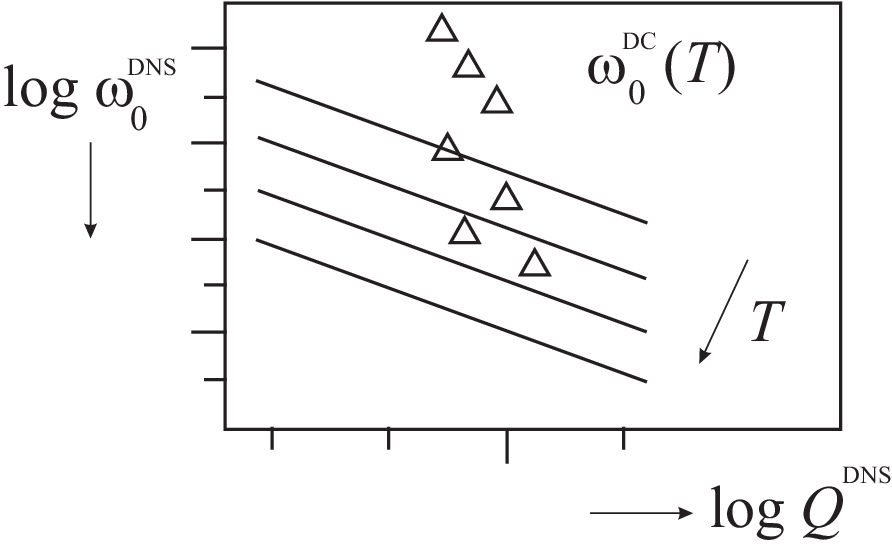}
\caption{Fictitious diagram for the Experimentum Crucis. The
maximum frequency values $\omega _0^{\text{DC}}(T)$ of dynamic heat capacity $%
C_p^{\prime \prime }(\omega ,T)$ for different given temperatures, $%
(\triangle )$, [located by the raster of isotherms, and using $\omega _0^{%
\text{DC}}=\omega _0^{\text{DNS}}$ from the FDT] in a log
frequency ($\log \omega _0^{\text{DNS}})-\log $ wave vector ($\log
Q^{\text{DNS}}$) diagram for that part of dynamic neutron
scattering which is informative about diffusive and corresponding
relaxation processes. DC the dynamic calorimetry, DNS the dynamic
neutron scattering. The $\log Q$ abscissa values of the DC
triangles $\Delta $ give qualitative measures of the lengths ($\xi
=2\pi /Q$) that can be associated with DC and can be compared
with the calorimetric characteristic lengths from (\ref{Eq.7.1}) and (\ref%
{Eq.7.2}), Fig.~6. Details see text. \label{fig7}}
\end{center}
\end{figure}

In Ref.~\cite{Tyagi2005}, DNS for polyvinylacetate PVAC is compared with the
peak of $\varepsilon ^{\prime \prime }(\omega )$, the dielectric loss. The
authors get a length of about 1 nanometer (from $Q=0.65$~\AA $^{-1}$). In
our experience with DC and pluralism, the trace of $\varepsilon ^{\prime
\prime }$ in the relaxation chart is usually located not too far from the $%
C_p^{\prime \prime }$ trace. The extrapolation of von Laue cooperativities
to $x=1$ (crossover) gives the order of one or a few particles, i.e. the
same length order. The Gibbs length would give cooperativities $N_{mt}$ of
order 100 - 1000, corresponding to lengths of about 2 - 5 nanometer, i.e.
(much) larger. This preliminary comparison actually indicates von Laue
thermodynamics.\newline
\newline

In a way, the defect model is directly supported by DNS for polyisoprene %
\cite{Arbe1998,Arbe2003} in a large $Q$ interval (0.1~\AA $^{-1}$...5.0~\AA $%
^{-1}$) and time interval ($\tau _{\text{KWW}}=0.03$~ns...200~ns). The
raster of Fig.~7 corresponds to a large ''homogeneous'' region (Levy
diffusion) for small $Q$ values (large lengths $\xi =2\pi /Q$), which is
continued by a ''heterogeneous'' region for large $Q$ values (smaller
lengths). The $\tau _{\text{KWW}}$ times there are above the raster
extrapolation. The crossover between the two regions is at $Q\approx 1.3$%
~\AA $^{-1}$, corresponding to $\xi \approx 5$~\AA , close to the maximum of
the static structure factor, $S(Q)$. This observation can be interpreted by
our defect model, Fig.~2c. The homogeneous region corresponds to the fractal
center of the defect. The large lengths can be explained by the analytical
continuation as used for the Levy diffusion to the Fischer modes
(Section~VI.F). The heterogeneous region corresponds to the diffusion step
through the cage door. A responsibility part of this region corresponds to
the preponderant component of the Levy sum, being, so to speak, an
''inhomogeneous'' contribution from the homogeneous Levy distribution. The
other part is from the heterogeneous molecular contributions in the
construction of the cage door up to large $Q$ (i.e. down to small lengths).
The observation of the homogeneous part supports one of our basis assumption
for the Levy statistics, that the large numbers needed for limit
distributions come from the many dynamic attempts in the partial systems,
and is not directly the number of particles participating at the cage door.%
\newline
\newline

Distinguish two results from the Experimentum Crucis. (A). Decision between
von Laue or Gibbs thermodynamics for minimal subsystems in liquids, and (B).
The values of the bulk characteristic lengths per se, in nanometers,
indicating the spatial extent of the entropy or temperature responsibility
in the defects.\newline
\newline

It should be tried to widen the reliable experimental ranges of dynamic
neutron scattering along the $mt$ dispersion zone to lower frequencies than
today really available (about several 10$^7$Hz) and of dynamic calorimetry
to higher frequencies (than the today 10$^4$Hz), so that both methods can
overlap for the same substances e.g. in the crossover region of the $mt$
(usually in the 10$^6$Hz range, cf. Ref.~\cite{DonthBuch2001}, p.222-223).
This would decide which alternative for the characteristic length would be
consistent with scattering. The typical glass transition crossover frequency
can be reached by dynamic neutron scattering in the next years, if e.g. the
impressive series of experiments by Dieter Richter's group (e.g. \cite%
{Arbe1998,Arbe2003,Frick1995}) is continued to lower frequencies. Reliable
dynamic calorimetry in the megahertz range, however, remains to be done \cite%
{SchickPC2006}. Dynamic photon scattering can probably be applied in the $%
\alpha $-process dispersion zone at low frequencies, a few frequency decades
above freezing-in around the glass temperature $T_g$. Experimental data from
dynamic calorimetry is available here \cite{Hempel2000,HuthDonthSchick}.%
\newline
\newline

The decision for von Laue would become a conclusive experimental argument
for Levy defects, if our treatment from defects to von Laue thermodynamics
has a robust reversal. This includes three main steps: (1) Freely
fluctuating subsystems should give a sound thermodynamic basis for the
components of a Levy sum with Levy exponent $\alpha \leq 1$
(Representativeness Theorem). (2). In particular, such equivalent subsystems
should give a thermodynamic basis for the way from Kohlrausch correlation
functions to the Levy sum (Theorem 1). (3). We need a sound explanation to
have a Levy instability in the liquid giving $\alpha <1$ (free volume
picture Fig.~1 and Theorem 2). I think we can explicitly find the robust
reversal in these steps.\newline
\newline

There are general consequences for thermodynamics, if the temperature
fluctuation at so small, one-nanometer lengths could be confirmed by the
Experimentum Crucis. Since we do not have a molecular spatio-temperature
field in this length range, i.e. no $T({\bf r},t)$ there, we cannot
illustrate temperature fluctuations by additional hydrodynamic arguments
using forces and fluxes \cite{Kondepudi1998}. Instead, $\delta T$ must be
included in the many possibilities of the molecular origin for the random
components of a Levy sum which is agitated by the high attempt rates of
dynamics. Instead of hydrodynamics, the Levy defect forms some Levy
diffusion. Temperature fluctuation is there a necessary part of the dynamic
molecular play in liquids.\newline
\newline

Teaching thermodynamics, we could leave the thermodynamic limit for the
definition of temperature. If $\delta T\neq 0$ would be confirmed, then the
Gibbs distribution comes from a ''static'' restriction: we start from a more
general thermodynamics for subsystems with $\delta T\neq 0$ and come to the
limit $N\rightarrow \infty $ with $\delta T=0$, e.g. by means of a heat
reservoir. The use of this restriction as a basis of the general definition
of temperature (as sometimes advocated \cite{Kittel1988}) would not longer
be reasonable. Temperature $T$ is essentially different from an energy $%
\Delta E$, irrespective of the frequent occurrence of an equation of type $%
T=\Delta E/k_B$. In liquids, such an equation comes from Planck's black body
radiation of the quantums $\Delta E=\hbar \omega $ in Nyquist's transmission
lines for a thermodynamic measurement, $\hbar \omega =k_BT$. In the FDT,
this equation connects temperature fluctuations and frequency fluctuations
for a specific response, as used above.\newline
\newline

In summary, an Experimentum Crucis which confirms the lengths calculated
from von Laue thermodynamics would overcome the nonfluctuation of
temperature $-$ i.e. brings a new (the old) definition of temperature in
liquids beyond (and before) the Gibbs definition with the latter's heat
reservoir without fluctuation. The consequence for liquids could be that
mechanics and thermodynamics cannot be separated into denominator over
nominator, ($E(p,q)/k_BT$), as for Gibbs. The ''statistical mechanics''
could be influenced by a thermodynamics with Levy defects in the liquid.
Then the observables, in particular the dynamic susceptibilities, cannot
exactly be calculated from mechanical computer simulation plus Gibbs
thermodynamics alone.\newline
\newline

\section{Conclusions}

Most of the qualitative attributes for the arrangement and properties of the
dispersion zones ($a,\alpha ,\beta ,\phi $) in the relaxation chart of the
dynamic glass transition (as a synonym of classical liquid dynamics) can be
explained by a weak underlying fluctuating free-volume pattern with strong
dynamic Levy defects. Eight Examples are:\newline
\newline

(1) The Levy defect explains the concentration of the main transition ($mt$)
dynamics into a relatively narrow dispersion zone with a half-width of about
($1/\alpha $) decades, with $\alpha \leq 1$ the Levy exponent. The spectral
densities for extensive variables are Levy distributions, their correlation
functions are stretched exponentials.\newline
\newline

(2) The spatial position of the Johari Goldstein $\beta $ process in the
pattern is the center of the Glarum Levy defect of the $\alpha $ process;
the continuous part of the $\beta $ process comes from the cooperative
periphery.\newline
\newline

(3) The location of the crossover region in the relaxation chart is defined
by virtual crossing of a quasi mechanical molecular barrier mechanisms
(Arrhenius process underlying the $\beta $ process) with the main $mt$
process being not an Arrhenius process.\newline
\newline

(4) The emerging von Laue thermodynamics allows an experimental
determination of the characteristic length for the main transition by bulk
dynamic calorimetry. This length is the size of minimal representative
subsystems and characterizes the size of the Glarum Levy defects (including
its preponderant component, its fractal center and its cooperative
periphery) and their mutual average distance in the pattern.\newline
\newline

(5) The preponderant component of the $mt$ Levy sum for the defect may
induce an extraordinary molecular event of dynamics: The diffusion step
through the cage door opened in the cage of equivalent next molecular
neighbors.\newline
\newline

(6) The non-Arrhenius behavior of the main transition $mt$ ($=a+\alpha $
process) is connected with the increase of cooperativity $N_{mt}$ at low
temperatures.\newline
\newline

(7) The position of the dispersion zone for the Fischer modes $\phi $
relative to the main transition in the relaxation chart is determined by the
diffusion-relevant Levy exponent $\alpha $ from the $mt$ Glarum Levy defects
without the need to refer to any thermodynamic collective structure
formation beyond the Levy distribution (no ''clusters''). The Fischer
speckles are the Levy defects of a Levy sum from a large number of minimal
subsystems.\newline
\newline

(8) At low temperatures (or mobilities), the main transition is probably
exhausted. The operational freedom in the periphery for large
cooperativities $N_{mt}$ becomes so small that spatially distributed, random
traces of foreign processes can take the control over from the main process.
This restricts the Levy exponent from below, $\alpha (mt)\gtrsim 0.4$.%
\newline
\newline

The way from the widely observed stretched exponential (Kohlrausch function)
to a Levy distribution for the spatial fluctuation pattern with preponderant
components as midst of the defects is explained by a succession of theorems
which are partly known before. Their original mathematical proofs are partly
modified and completed by physical arguments and comments. Two things are
additionally needed for this way: Mathematically, (1) a sum of Levy sum
components and (2) an explanation of small Levy exponents $\alpha <1$. This
corresponds physically (ad 1) to a sum of spatially separable statistically
independent subsystems or partial systems and (ad 2) to a Levy instability
from a local breakthrough of mobility. The latter is characterized by more
free volume in the center as well as a spectrum of relaxation in the
cooperative periphery of the defect. This ''inside explanation'' is
confirmed by an ''outside'' one from the Representativeness Theorem for
thermodynamics of liquids, operating with a sum of representative minimal
subsystems. An Experimentum Crucis for the whole argumentation is suggested:
the comparison of characteristic lengths from thermodynamics (dynamic
calorimetry) with lengths from dynamic neutron (or photon) scattering.%
\newline
\newline

The Levy distribution gets a ''shaping power of statistics'' if it is
imbedded in certain additional circumstances. For the dynamics in classical
liquids, e.g., the following is sufficient: (a). Many equivalent independent
spatially separable partial systems for its establishment as a limit
distribution, (b). An instability whose stabilization by the limit
distribution is different from Gauss (exponent $\alpha <2$, in particular $%
\alpha <1$ for the existence of a preponderant component in the Levy sum),
and (c). A general time (mobility)-length scaling of relevant fluctuation
modes to get a fluctuating space pattern with defects. The latter are thus
considered as the results of the shaping power.\newline
\newline

Acknowledgment. The author thanks Profs. Thomas Thurn-Albrecht (Halle),
Dieter Richter (J\"ulich), Christoph Schick (Rostock), Gyan P. Johari
(Hamilton), and Dr. Klaus Schr\"oter (Halle) for pleasant discussions since
2001.

\end{document}